%% file: Part_II.tex
\documentclass[journal]{IEEEtran}
\input{packages}

\input{commands}

\input{acronyms}
\begin{document}
\bstctlcite{IEEEexample:BSTcontrol}
\title{Radio Localization and Sensing -- Part II: State-of-the-art and Challenges}
\specialpapernotice{(Invited Paper)}
\input{authors}
\setlength{\abovedisplayskip}{4pt}
\setlength{\belowdisplayskip}{4pt}
\maketitle
\begin{abstract}
This letter is part of a two-letter tutorial on radio localization and sensing, with a focus on mobile radio systems, i.e., 5G and beyond. Building on Part I, which focused on the fundamentals, here we go deeper into the state-of-the-art advances, as well as 6G, 
covering enablers and challenges related to  modeling, coverage, and accuracy.
\end{abstract}
\vspace{-5mm}
\section{Introduction}
\begin{figure}
    \centering
    \includegraphics[width=0.9\linewidth]{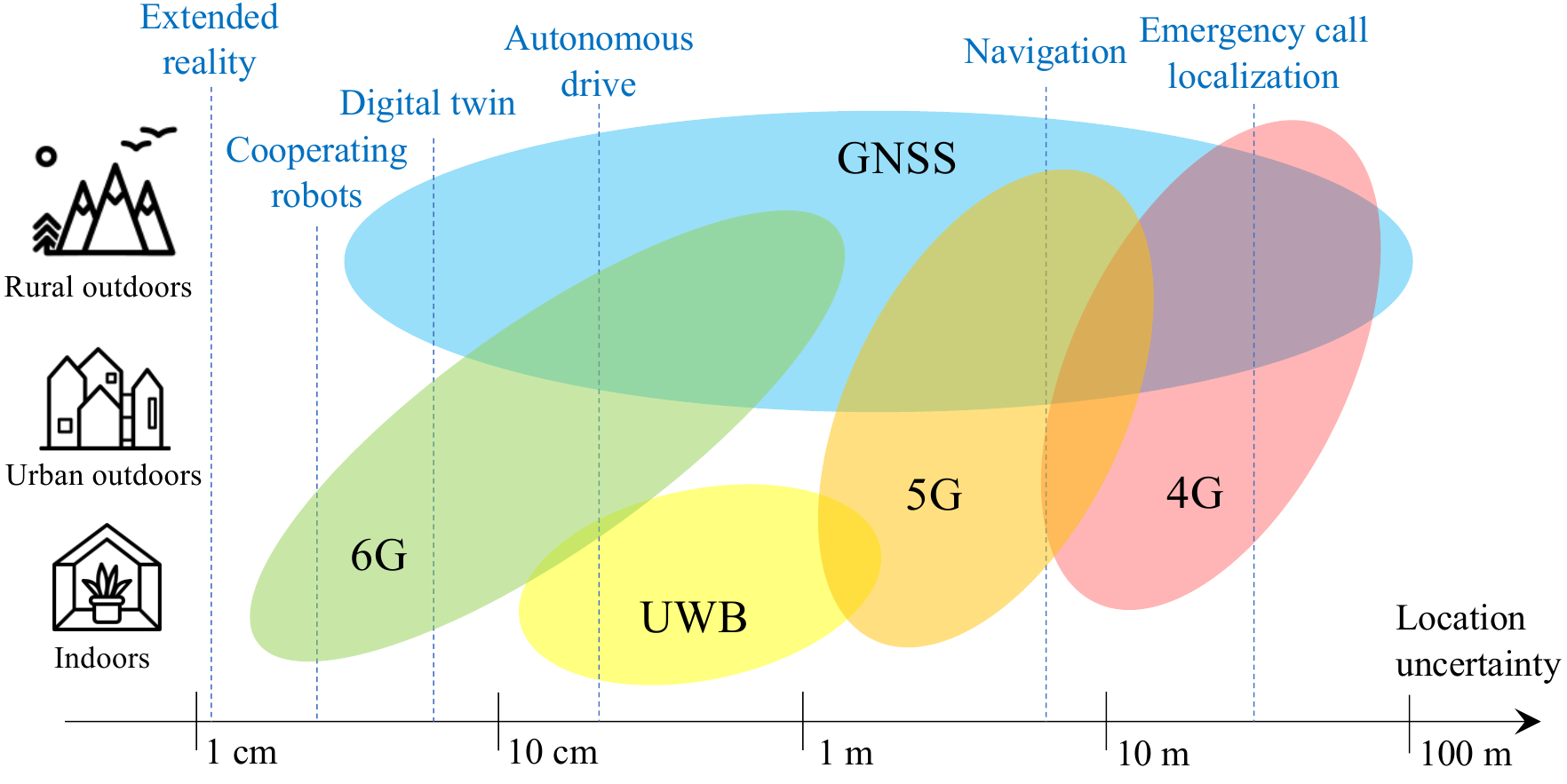}\vspace{-3mm}
    \caption{Localization accuracies for different radio technologies in different environments. Selected use case requirements are shown in blue. \ac{GNSS} positioning has a wide range of accuracies, depending on whether terrestrial support or carrier phase measurement are available. Inspired by~\cite[Fig.~8]{SurveyCellularRadioLocalization--Rosado_others_G.Seco-Granados}.}
    \label{fig:6Grevised} \vspace{-4mm}
\end{figure}

\PARstart{H}{igh-performance} localization and sensing requires combination of sufficient \emph{coverage} in terms of infrastructure, \emph{resolution} of different signal paths, and finally estimation \emph{accuracy} of the geometric parameters of each resolvable path. Fig.~\ref{fig:6Grevised} shows different radio technologies, including \acp{GNSS} and \ac{UWB}~\cite{win2009history}, in terms of their achievable accuracy for different environments, as well as requirements of selected use cases. With \ac{GNSS} being limited to outdoors and \ac{UWB} to short-range applications, 4G, 5G, and 6G provide a bridge. 
Arguably the main advances towards 5G were the introduction of massive MIMO and the use of mmWave spectrum~\cite{andrews2014will}. Massive MIMO provides high angular resolution
~\cite{wen2019survey}, but in sub-6 GHz bands still exhibits limited positioning performance, due to the relatively small delay resolution and multipath rich propagation channels. In contrast, mmWave communication (30--300 GHz, though 5G utilizes only lower mmWave bands around 24--53 GHz) has provided unique opportunities for localization, especially when combined with massive arrays, typically analog or hybrid arrays,~\cite{PositionLocforFuturisticCommunications--O.Kanhere_T.Rappaport}. In 6G, which is expected to utilize upper mmWave bands (100--300 GHz), with even larger available spectrum, cm-level accuracy is  attainable~\cite{chen2021tutorial}. 

Let us first review  the reasons \emph{why} 5G and 6G mmWave systems are expected to provide such exceptional performance. 
First, the use of higher carrier frequencies leads to a more benign propagation channel 
that is more closely related to the geometry, with a relatively small number of propagation clusters 
\cite{rappaport2017overview,peng2020channel}. 
Second, higher carrier frequencies allow transmission of larger contiguous bandwidths (up to 400 MHz in 5G), providing much better delay resolution than at lower frequencies, where bandwidths are limited to a few tens of MHz. Third, for a fixed physical footprint, a larger number of antenna elements can be fit at the \ac{UE} and the \ac{BS} sides, enabling fine beamforming and providing enhanced angular resolution. 
These benefits come at the cost of reduced coverage, sensitivity to blockages and hardware impairments, and higher power consumption than sub-6 GHz deployments.

\begin{figure}
    \centering
    \includegraphics[width=1.0\linewidth]{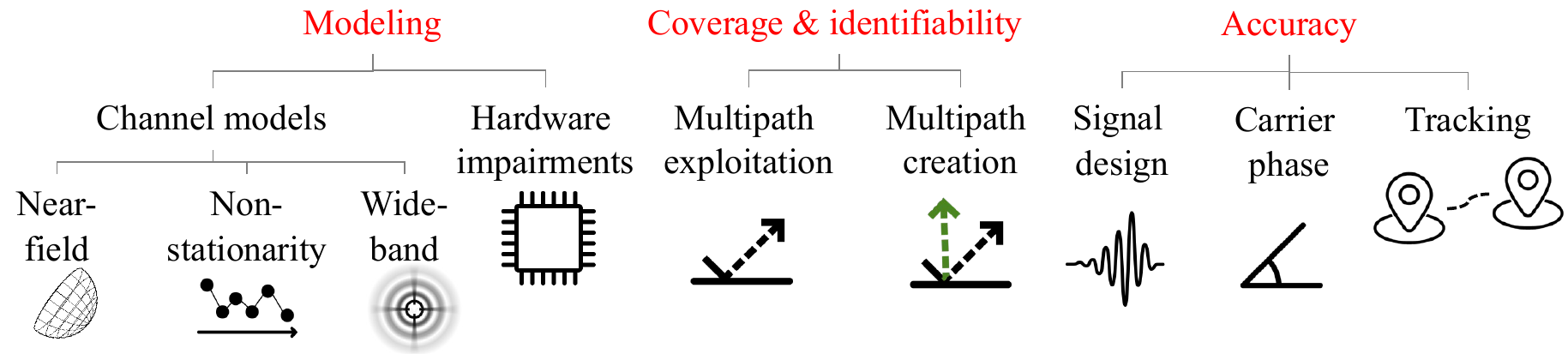} \vspace{-4mm}
    \caption{{Advances in models, improvements in coverage and in accuracy for 6G localization and sensing.}}
    \label{fig:PartIIOverview}\vspace{-6mm}
\end{figure}

In parallel, there have been several important developments in communication systems, which provide complementary opportunities for localization and  sensing. First, there is the increased importance of \ac{ISAC}, reusing hardware and possibly signals for both sensing 
and communication functions, with mutual benefits~\cite{liu2022integrated}. Second, the deployment of \acp{RIS}, which are largely passive devices that can modify the propagation environment for improving various \acp{KPI}~{\cite{strinati2021reconfigurable,basar2019wireless}}. Third, the introduction of other MIMO-based technologies, such as cell-free massive MIMO~\cite{demir2021foundations}, holographic MIMO~\cite{huang2020holographic}, and \ac{XL-MIMO}~\cite{bjornson2019massive}. Fourth, there are algorithmic advances, such as the use of \ac{AI} to tackle certain classes of problems that are hard to solve with model-based methods~\cite{o2017introduction}, as well as novel localization and mapping methods~\cite{kim2022pmbm}, analysis tools~\cite{fortunati2017performance}, and signal designs~\cite{keskin2022optimal}. 

In this letter, we will cover several of these advances,  
broken down 
into modeling, enablers for improving coverage and identifiability, and enablers for improving accuracy{, as shown in Fig.~\ref{fig:PartIIOverview}}.

\section{Advances in Modeling} \label{sec:advancedmodels}
The models from Part I are only valid for a limited operating range. When we go to extremes in terms of bandwidths or array sizes, additional effects should be considered. Moreover, we have ignored the impact of hardware impairments, which will degrade  localization and sensing performance. 
\vspace{-3mm}
\subsection{Radio Channel Models}

We recall the channel model from Part I, between a \ac{Tx} with location $\bm{x}_{\text{tx}}$ and a \ac{Rx} with location $\bm{x}_{\text{rx}}$ at frequency $n\Delta_f$ (or subcarrier $n$) and symbol $k$
\begin{align}
    \bm{H}_{n,k} = \sum_{l=1}^{L} \alpha_l \bm{a}_{\text{rx}}(\bm{\theta}_{l})\bm{a}_{\text{tx}}^\top(\bm{\phi}_{l})e^{-\jmath 2 \pi n  \Delta_f \tau_{l}}  e^{\jmath 2 \pi k T_s  \nu_l}. \label{eq:ChannelGeneric}
\end{align}
Suppose path $l>1$ corresponds to an \ac{IP}  $\bm{x}_{\text{inc},l}\in \mathbb{R}^3$. The steering vector $\bm{a}_{\text{rx}}(\bm{\theta}_l)$ (similarly for $\bm{a}_{\text{tx}}(\bm{\phi}_l)$) has entries $   [\bm{a}_{\text{rx}}(\bm{\theta}_l)]_p = \exp({\jmath (\bm{x}_{\text{rx},p}-\bm{x}_{\text{rx}})^\top \bm{u}(\bm{\theta}_l) {2 \pi}/{\lambda} )})$, 
where $\bm{x}_{\text{rx},p}-\bm{x}_{\text{rx}}$ is the location of the $p$-th antenna element, and $\bm{u}(\bm{\theta}_l)=(\bm{x}_{\text{inc},l}-\bm{x}_{\text{rx}})/\Vert \bm{x}_{\text{inc},l}-\bm{x}_{\text{rx}}\Vert $, all expressed in the \ac{Rx} frame of reference. 

\subsubsection{Near-Field Propagation}
The previous model is in fact a limiting form of a more general model that accounts for wavefront curvature (often called the near-field model)~\cite{friedlander2019localization}, $[\bm{a}_{\text{rx}}(\bm{x}_{\text{inc},l})]_p=\exp({-\jmath 2 \pi (d_p-d_{\text{ref}}) /\lambda })$, 
where $d_{\text{ref}}=\Vert \bm{x}_{\text{inc},l}-\bm{x}_{\text{rx}}\Vert $ is the distance between the source and array's phase reference, while $d_p=\Vert \bm{x}_{\text{inc},l}-\bm{x}_{\text{rx},p}\Vert $ is the distance between the source and the array's $p$-th element. 
Exploiting this wavefront curvature leads to new opportunities to improve accuracy~\cite{guidi2021radio},  coverage~\cite{rahal2021ris}, and  signal designs~\cite{rinchi2022compressive}. 

\subsubsection{Channel Non-stationarity}
 Non-stationarity refers to the variation of the channel gain across an array and was first observed in \ac{XL-MIMO} systems~\cite{de2020non}. We focus on the \ac{LoS} path in a localization context, where the usual gain expression is $|\alpha_1|^2=(\lambda^2/(4 \pi)^2)G_{\text{rx}}(\bm{\theta}_{1})G_{\text{tx}}(\bm{\phi}_{1})/ \Vert\bm{x}_{\text{tx}}-\bm{x}_{\text{rx}} \Vert^2$~\cite[Eq.~(3)]{WymSec22LetterPartI}. Under XL-MIMO conditions, the channel gain depends on the \ac{Tx} antenna index $q$ and the \ac{Rx} antennas index $p$: 
\begin{align}
    |\alpha_{1,p,q}|^2=\frac{\lambda^2}{(4 \pi)^2} \frac{G_{\text{rx}}(\bm{\theta}_{1,p,q})G_{\text{tx}}(\bm{\phi}_{1,p,q})}{\Vert\bm{x}_{\text{tx},q}-\bm{x}_{\text{rx},p} \Vert^2}.
\end{align}
The effect can in principle be decoupled from the wavefront curvature but is often considered jointly. 
If the channel gain varies significantly as a function of $p$ or $q$, conventional localization and sensing methods must be reconsidered, especially  the channel parameter estimation routines~\cite{lu2021communicating}. 

\begin{figure}
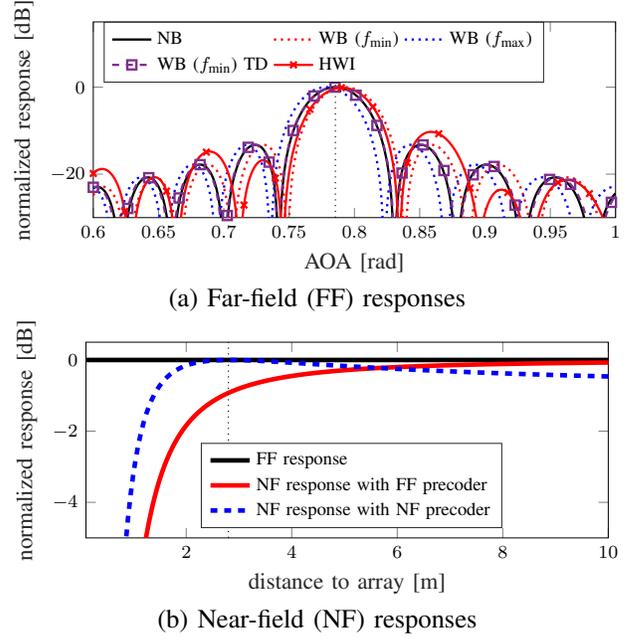

\begin{minipage}{0.98\linewidth}
  \centering
  \include{figures/FiguresPartII/letterPartII-ModelsA}
    \vspace{-0.5cm}
    \centerline{(a) Far-field (FF) responses} 
\end{minipage}
\hfill
\begin{minipage}{0.98\linewidth}
  \centering
    \include{figures/FiguresPartII/letterPartII-ModelsB}
    \vspace{-0.5cm}
    \centerline{(b) Near-field (NF) responses}
\end{minipage}
\caption{Array responses in narrowband (NB) and wideband (WB) for a system at 28 GHz with 400 MHz bandwidth, using a 64-element linear array receiver and a single-antenna transmitter at 2.8 m  and \ac{AoA} of $\pi/4$. Beam squint is visible in (a) when phase control is used:  the responses at lowest $f_{\min}$ and highest $f_{\max}$ frequency become separate from the main beam direction. Beam squint  disappears under time delay (TD) control. Hardware impairments (HWI) in the form of small antenna displacements affect the response.  In (b) when ignoring path loss, the FF response is independent of the distance to the array, while 
the NF response leads to power penalties at short distances when a FF precoder is used. A NF precoder can focus energy at a target distance.  }
\label{fig:Models}
\vspace{-5mm}
\end{figure}

\subsubsection{Wideband Effects}
A final effect we consider pertains to the bandwidth, and is known as beam squint, which is an unfocusing or shifting of beams in an array~\cite{myers2021infocus}. Previously, we have used $\lambda$ to denote the wavelength, which in fact represents the wavelength at the carrier, $\lambda=c/f_c$. If the signal bandwidth $W$ is significant with respect to the carrier, say $W/f_c > 0.1$, then the wavelength becomes frequency-dependent: $\lambda_n = c/(f_c+n \Delta_f)$, where $n$ is the subcarrier index and $\Delta_f$ the subcarrier spacing. This wideband effect is especially pronounced for large arrays and should thus be considered together with near-field and non-stationarity effects. When localization or sensing ignores beam squint, significant performance degradation can occur. Hence, dedicated channel parameter estimation routines must be applied~\cite{wang2019beam}. Alternatively, frequency-dependent precoders and combiners, implemented with true time delays can remove beam squint.
\vspace{-4mm}
\subsection{Hardware Impairments}
Reaching extreme performance requires extreme calibration and puts extreme demands on the communication hardware.  Hardware impairments in transceivers can be broken down into synchronization errors (phase noise, clock frequency offsets or drifts, timing errors), array errors (mutual coupling between antenna elements, unknown element responses, array element displacements), and other effects (e.g., power amplifier nonlinearity and quantization) ~\cite{mohammadian2021rf}. Synchronization errors are generally time-varying and must thus be tracked and mitigated. Since most localization and sensing methods rely on very precise synchronization as well as measuring phase across time, frequency, and space, synchronization methods must be not only powerful, but also take care of not removing valuable Doppler and phase information. 
On the other hand, array errors are largely static errors, due to the lack of proper calibration. Hence, these can be learned and mitigated over time. As many of the impairments are nonlinear and dispersive, \ac{AI}-based methods are well-suited to mitigate them. 
\vspace{-3mm}
\subsection{Case Study and Analysis Tools}
In Fig.~\ref{fig:Models}, we show the antenna responses of a linear array as a function of the azimuth angle and distance, for a 28 GHz system with 64 antennas and 400 MHz bandwidth. The figure shows that the considered effects are all non-negligible, even for these relatively modest parameters. To understand which effect becomes important for a specific localization or sensing problem, the \ac{MCRB} is a powerful tool~\cite{fortunati2017performance}, as it can lower-bound the performance of the mismatched maximum likelihood estimator~\cite{chen2022mcrb}.


\vspace{-3mm}

\section{Improving Coverage and Identifiability}
\begin{figure}
    \centering
    \includegraphics[width=0.9\columnwidth]{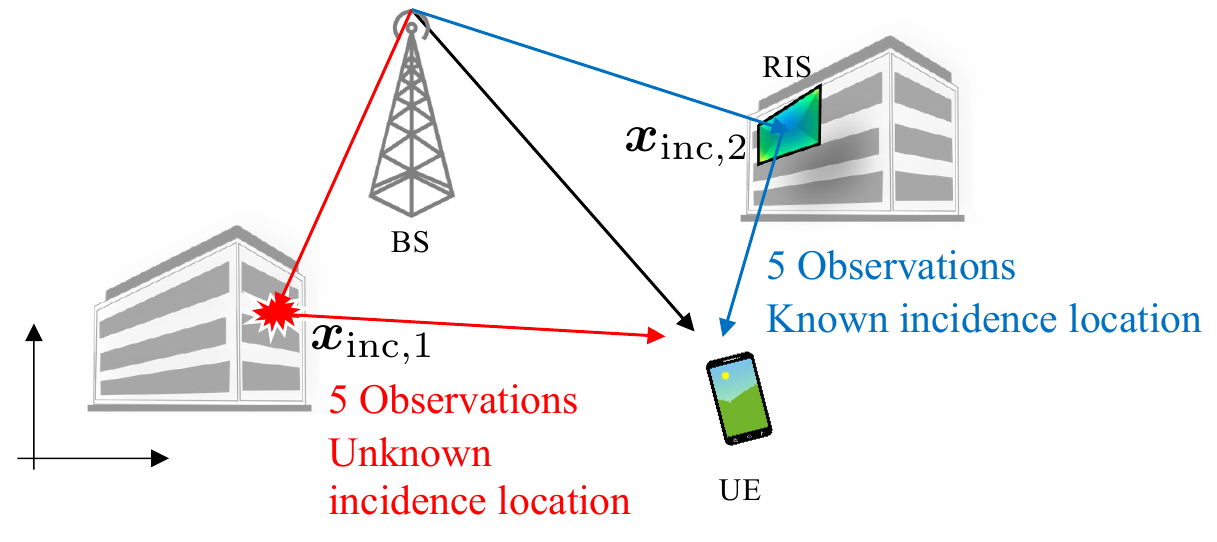}
    \caption{Exploitation of uncontrolled multipath and generation of controlled multipath by RIS improves localization coverage, since each path provides 5 geometric observations (4 angles, 1 delay).}
    \label{fig:MPE}\vspace{-5mm}
\end{figure}

Coverage and identifiability are limited by the \ac{BS} deployment, but can be improved by multipath exploitation and control, see  Fig.~\ref{fig:MPE}.

\vspace{-6mm}
\subsection{Multipath Exploitation} \label{sec:MPE}
{
Multipath exploitation is a generalization of  both localization and bistatic sensing of passive objects\footnote{{This is the common form of multipath exploitation. Alternative approaches rely on \ac{RIS} or on prior map information, for both monostatic and bistatic configurations.}}, benefiting from the \ac{NLoS} part of the channel, rather than considering it as a disturbance~\cite{HighAccuracyLocalizationforAssistedLiving--Witrisal_others,wielandt2017indoor}.}
By considering each propagation path in \eqref{eq:ChannelGeneric} as an object, generating a specific \ac{ToA}, \ac{AoA}, \ac{AoD}, and Doppler, we can infer the 3D location of \acp{IP} $\bm{x}_l$, e.g., via the relation $\tau_{l} = (\Vert \bm{x}_{\text{tx}}-\bm{x}_{\text{inc},l}\Vert + \Vert \bm{x}_{\text{rx}}-\bm{x}_{\text{inc},l}\Vert)/c + B$. 
Since each multipath component has three spatial degrees of freedom (i.e., $\bm{x}_{\text{inc},l}\in \mathbb{R}^3$) and a 5-dimensional measurement (ToA, AoA and AoD in azimuth and elevation, if we ignore Doppler), multipath directly contributes to improving localization performance~\cite{nazari2022mmwave}.  
In Table \ref{tab:CoverageAnalysis}, we provide an overview of minimal configurations needed to solve the 3D localization problem (and in the MIMO cases, the 3D orientation problem). Observe that conventionally 4 BSs are needed, while a single BS suffices under multipath exploitation~\cite{mendrzik2018harnessing}. 
Finally, note that we have ignored the effect of (i) diffuse multipath (which can be included as part of the noise, though with location-dependent statistics) and (ii) multi-bounce effects, in which case there are several \acp{IP} per path, where the AoD relates to the first \ac{IP} and the AoA relates to the last \ac{IP}, while the ToA depends on the end-to-end propagation distance.

\vspace{-4mm}

\subsection{Multipath Creation by RIS}
We can also \emph{create} multipath with the aid of RIS, to improve localization~\cite{keykhosravi2022ris} or sensing~\cite{buzzi2022foundations}. An \ac{RIS} leads to an extra term in the channel \eqref{eq:ChannelGeneric} of the form 
\begin{align}
    \bm{H}^{\text{ris}}_{n,k} = \alpha^{\text{ris}}_{k} \bm{a}_{\text{rx}}(\bm{\theta}^{\text{ris}})\bm{a}_{\text{tx}}^\top(\bm{\phi}^{\text{ris}})e^{-\jmath 2 \pi n  \Delta_f \tau^{\text{ris}}}  e^{\jmath 2 \pi k T_s  \nu^{\text{ris}}}. \label{eq:RISChannel}
\end{align}
The main difference over \eqref{eq:ChannelGeneric} lies in the controllable nature of $\alpha^{\text{ris}}_{k} = \alpha^{\text{tx-ris}}\alpha^{\text{ris-rx}} \bm{a}_{\text{ris}}^\top({\bm{\phi}}^{\text{ris-rx}}) \bm{\Omega}_k \bm{a}_{\text{ris}}({\bm{\theta}}^{\text{tx-ris}})$, 
%
in which $\alpha^{\text{tx-ris}}$ is the complex gain from Tx to RIS, $\alpha^{\text{ris-rx}}$ is the complex gain from RIS to Rx, $\bm{a}_{\text{ris}}(\cdot)$ is the RIS response vector, assuming far-field operation, 
as a function of the AoA from the Tx  ${\bm{\theta}}^{\text{tx-ris}}$ and the AoD to the Rx ${\bm{\phi}}^{\text{ris-rx}}$. If  \ac{Tx} is a \ac{BS}, the \ac{AoD} $\bm{\phi}^{\text{ris}}$ and \ac{AoA} ${\bm{\theta}}^{\text{tx-ris}}$ are known and do not need to be estimated. 
The RIS configuration is set by $\bm{\Omega}_k$, a diagonal matrix with entries $\omega_{m,k} \in \mathcal{W} \subset \mathbb{C}$, where $\mathcal{W}$ is a predetermined set of RIS element configurations that depend on the technology{, which may introduce undesired latency}. Without amplification,  $|\omega_{m,k}|\le 1$, so that $\bm{a}_{\text{ris}}^\top({\bm{\phi}}^{\text{ris-rx}}) \bm{\Omega}_k \bm{a}_{\text{ris}}({\bm{\theta}}^{\text{tx-ris}}) \le M$, where $M$ is the number of RIS elements.   
An RIS with known location thus provides a high-dimensional geometric observation (ToA, a 2D angle at the UE, a 2D angle at the RIS and possibly a Doppler) without any additional unknowns. Hence, the RIS acts as a secondary synchronized BS with a phased array, sending the same signal as the real BS. Consequently, as shown in the  last row of Table \ref{tab:CoverageAnalysis}, a single BS and a single RIS are sufficient to localize a user. The introduction of an RIS does not affect resolvability, as the temporal encoding $\bm{\Omega}_k$ allows it to be separated from the uncontrolled multipath~\cite{keykhosravi2022ris}. 




\newcolumntype{P}[1]{>{\centering\arraybackslash}m{#1}}
\begin{table}[]
    \centering
    \resizebox{\columnwidth}{!} {
    \begin{tabular}{|P{2.5cm}|P{2cm}|P{2cm}|P{2cm}|P{2cm}|}
    \hline
     & \textbf{Angle-only SISO} & \textbf{Angle-only MIMO} & \textbf{Angle \& delay SISO} & \textbf{Angle \& delay MIMO} \\
    \hline
    \hline
    \textbf{BS only} & not applicable & 2 BS & 4 BS &  2 BS \\
    \hline
    \textbf{BS + multipath} & not applicable & 2 BS & 4 BS &  1 BS, 1 IP~\cite{nazari2022mmwave} \\
    \hline
    \textbf{BS + multipath, no LOS} & not applicable & not identifiable & not identifiable & 1 BS, 4 IP~\cite{mendrzik2018harnessing} \\
    \hline
    \textbf{BS + RIS} & 1 BS, 2 RIS & 1 BS, 1 RIS & 1 BS, 1 RIS\cite{keykhosravi2022ris}  & 1 BS, 1 RIS  \\
    \hline
    \end{tabular}
    }
     \vspace{0.1mm}
    \caption{Minimal configurations needed to solve the static 3D localization problem without a priori knowledge of \acp{IP} locations. MIMO configurations also consider an unknown 3D orientation. Configurations that rely on delay measurements require estimating the user clock bias. }
    \label{tab:CoverageAnalysis}\vspace{-8mm}
\end{table}

\vspace{-3mm}
\section{Improving Accuracy}

Coverage improvements will directly lead to improved accuracy. However, there are additional measures one can take to improve accuracy for any given (optimized \cite{albanese2022loko}) deployment. 
\vspace{-5mm}
\subsection{Signal Design}
\begin{figure}
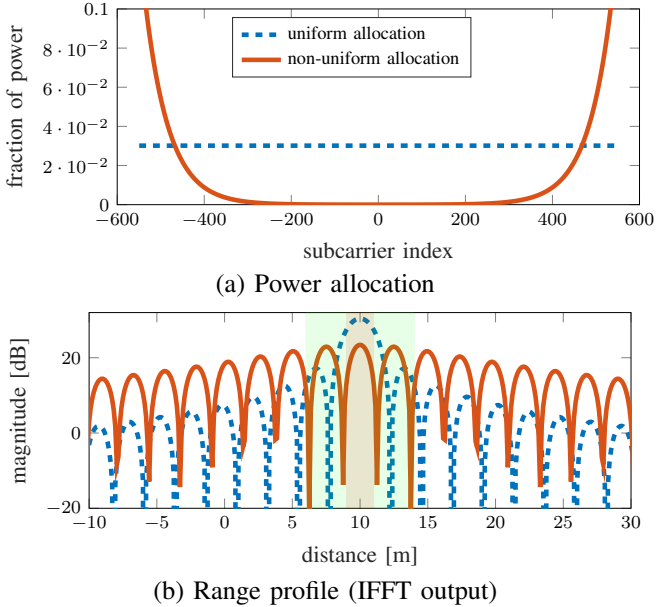

\begin{minipage}{0.98\linewidth}
  \centering
    \include{figures/FiguresPartII/letterPartII-power}
    \vspace{-0.5cm}
    \centerline{(a) Power allocation} \medskip
\end{minipage}
\hfill
\begin{minipage}{0.98\linewidth}
  \centering
    \include{figures/FiguresPartII/letterPartII-AF}
    \vspace{-0.5cm}
    \centerline{(b) Range profile (IFFT output)}
\end{minipage}
\caption{Signal optimization to improve distance estimation, based on OFDM power allocation in (a) with 132 MHz bandwidth for a user at around 10 m. The green and red regions in (b) represent the a priori information.}
\label{fig:power-allocation}\vspace{-5mm}
\end{figure}

Given the deployment, 
the system can still optimize the signals sent over the channel, both in terms of time-frequency allocation, as well as through the used transmitter precoders, which we all described by $\bm{f}_{n,k}$ in \cite[Eq.~(2)]{WymSec22LetterPartI}. 
{When there is no a priori location knowledge regarding users/objects, broadcast signals are used, e.g., time-frequency comb pilot signals and an exhaustive sweep of directional beams for positioning and bistatic sensing, or communication-optimal designs with random data for monostatic sensing.} 
As soon as partial information becomes available, it can be used to optimize the signals{, applicable for both pilot \cite{keskin2022optimal} and data signals \cite{keskin2021limited}.} Signal designs can be based on the \acs{CRB}, with the explicit goal of maximizing accuracy~\cite{keskin2022optimal}. Given a user with a state uncertainty range $\bm{s} \in \mathcal{S}$ and a set of design parameters $\bm{d} \in \mathcal{D}$, then a localization-optimal design is of the form
\begin{subequations}
\begin{align}
    \mathrm{minimize}_{\bm{d\in \mathcal{D}}} \quad & \max_{\bm{s}\in \mathcal{S}} \, \text{PEB}(\bm{s}|\bm{d}) \\
    \text{s.t.} \quad &  \mathrm{identifiable}(\bm{s}|\bm{d}), \forall \bm{s}\in \mathcal{S},
\end{align}
\end{subequations}
where the \ac{PEB} was defined in Part I and the constraint `$\mathrm{identifiable}(\bm{s}|\bm{d})$' ensures that the design $\bm{d}$ does not lead to ambiguities in the localization estimates at $\bm{s}$. 
To exemplify the design problem, consider the power allocation problem across subcarriers in an \ac{OFDM} system~\cite{driusso2014performance,montalban2013power}, for performing distance estimation, shown in Fig.\,\ref{fig:power-allocation}-(a). The range profiles (correlation output used for maximum likelihood estimation) are shown in Fig.\,\ref{fig:power-allocation}-(b).
The uniform allocation (in blue) has a broad main peak (about $4.4~\text{m}$), but around 13 dB suppression of sidelobes. 
The red power allocation, which emphasizes the outer subcarriers is thus near-optimal from a PEB perspective with a  narrower main peak (about $2.3~\text{m}$, leading to a PEB reduction of about 60\%), but with several strong sidelobes.  Hence, the allocation should account for prior information (green and red areas in Fig.\,\ref{fig:power-allocation}-(b)) to meet the identifiability constraint.  
\vspace{-4mm}
\subsection{Carrier Phase Based Localization}
We argued in Part I that $\alpha_{l}$ in \eqref{eq:ChannelGeneric} should not directly be used for localization or sensing. In fact, this is only partially true. The phase of $\alpha_{l}$ (say, $\psi_l = \angle \alpha_l $) can be expressed as 
\begin{align}
    \psi_l = - \frac{2 \pi  (\Vert \bm{x}_{\text{tx}}-\bm{x}_{\text{inc},l}\Vert + \Vert \bm{x}_{\text{rx}}-\bm{x}_{\text{inc},l}\Vert)}{\lambda}   + \psi_{\text{tx}} +  \psi_{\text{rx}} + \tilde{\psi}_l, \notag 
\end{align}
where $\psi_{\text{tx}}$ is a phase due to the transmission chain at the \ac{Tx} and $\psi_{\text{rx}}$ is a phase due to the reception chain at the \ac{Rx}, $\tilde{\psi}_l$ is a phase induced by the reflection (for $l\neq 1$), 
and the remainder of $\psi_l$ is a function of the distance and the carrier frequency. Note that the phase $\psi_l$ is subject to a $2 \pi z$ ambiguity, $z\in \mathbb{Z}$.
Carrier phase information has two complementary benefits:
\begin{itemize}[noitemsep,topsep=0pt]
    \item \emph{Improved accuracy:} For example, in \ac{DL} localization with several \acp{BS}, $\psi_{\text{tx}}$ can be known from calibration for each \ac{BS}, so that when the correct ambiguity value can determined and the relation 
    $\Vert\bm{x}_{\text{tx}}-\bm{x}_{\text{rx}}\Vert= \lambda({-\psi_1+\psi_{\text{tx}} +  \psi_{\text{rx}}})/({2 \pi })$ can be exploited, then the \ac{UE} can be localized with an accuracy of a fraction of the wavelength $\lambda$ ~\cite{fan2021carrier}.
\item \emph{Improved resolution:} For example, in bistatic sensing, where several widely distributed \acp{Tx} and \acp{Rx} are phase-synchronized, an object with location $\bm{x}_{\text{inc},l}$ can be resolved with $\lambda$-level  resolution~\cite{haimovich2007mimo}, providing a way to obtain high resolution with few antennas and limited bandwidth, e.g., at sub-6 GHz frequencies.
\end{itemize}
\vspace{-4mm}
\subsection{Inference over Time}
Localization and sensing accuracy can also be improved by tracking users and objects over time {or by fusion with other sensors \cite{guo2020hybrid}, provided statistical models for the uncertainties are available}. Bayesian filtering methods such as the extended Kalman filter are well-suited to this~\cite{koivisto2017joint}. When \ac{UE} tracking is combined with multipath exploitation, this is known as \ac{SLAM},\footnote{In particular, as `channel SLAM' or `radio SLAM', as well as `5G SLAM' and `mmWave SLAM'. When the objects also move, the problems are referred to as tracking and \ac{SLAT}} while in a pure sensing context it is known as mapping {or tracking, which inherently exploits multipath}. These problems are often tackled after the range, angle, and Doppler measurements for the resolved paths are obtained. 
Both SLAM and mapping/tracking are inherently challenging because (i) the number of objects is a priori unknown, objects may not give rise to measurements at each time, while measurements may be due to clutter; (ii) the association between measurements and objects is unknown. Both challenges are elegantly addressed by either \ac{BP}~\cite{leitinger2019belief} or \ac{RFS} theory~\cite{kim2022pmbm}, while performance bounds can be obtained using the posterior \acs{CRB}~\cite{ristic2007comments}.

\vspace{-3mm}
\section{Outlook towards 6G}
With several initiatives worldwide underway to develop the basic building blocks of 6G, several of the models and methods covered in this letter will need to be used. The main trends we see towards 6G are the following:
\begin{enumerate}
    \item \emph{From position to orientation and velocity:} Conventional positioning has only considered the 3D location of the \ac{UE}. 6G will certainly include the 3D \ac{UE} orientation, as users will have arrays. In addition, to ensure high-SNR operation with long integration types, Doppler and micro-Doppler processing will be beneficial, so that velocity of users and objects can be inferred. 
    \item \emph{Utilizing a variety of carrier frequencies:} Much emphasis has been placed on lower and upper mmWave frequencies as main enablers for accurate localization and sensing, but we should not ignore the sub-10 GHz carriers, which provide better energy efficiency and coverage, while resolution can come from wide aperture deployments. 
    \item \emph{The rise of AI:} This letter has very much emphasized model-based signal processing. However, in the presence of model uncertainties (as described in Section \ref{sec:advancedmodels}) or algorithm deficiencies, data-driven methods can lead to disruptive and powerful designs and algorithms. With increased resolution come opportunities to infer finer details and properties of objects, people, and materials, similar to image processing. 
    \item \emph{ISAC:} 
    The use of common hardware and possibly common waveforms for communication and sensing (including localization) will provide important cross-functional benefits, such as reduced overhead beam alignment or radar sensing without dedicated radio emissions. 
    \item \emph{New antenna structures:} 
    Both at the user and infrastructure sides, many different antenna and deployment alternatives are being considered for 6G, all of which have implications for localization and sensing. Phase coherence among distributed infrastructure nodes will unlock the ultimate performance, though is very challenging to achieve in practice. 
\end{enumerate}
While by no means exhaustive, this short list, along with the models, methods, and challenges highlighted in these letters will hopefully spark new research questions, lead to novel methodologies, and ultimately help achieve the extreme performance demanded by 6G use cases. Based on several decades of experience, we can confidently say that the golden age of radio localization and sensing has only just begun. 

\vspace{-5mm}
\bibliographystyle{IEEEtran}
\bibliography{IEEEabrv,references} 
\end{document}

%% file: packages.tex
\usepackage[utf8]{inputenc}
\usepackage{bm}
\usepackage{epsfig,amsfonts,amsbsy,bm,mathrsfs}
\usepackage{amssymb,amsmath,amsthm,latexsym,amscd,amsfonts}
\usepackage{lettrine}
\usepackage{authblk}
\usepackage{graphics}
\usepackage{graphicx}
\usepackage{psfrag,float}
\usepackage{pstricks}
\usepackage{pst-plot}
\usepackage{cite}
\usepackage{balance}
\usepackage{epsfig}
\usepackage{epstopdf}
\usepackage{bbm}
\usepackage{enumitem}
\usepackage{dsfont}
\usepackage{setspace}
\usepackage{float}
\usepackage{relsize}
\usepackage{comment}
\usepackage{subfigure} 
\usepackage{algorithm}
\usepackage[noend]{algpseudocode}
\usepackage[nolist]{acronym}
\usepackage{tabularx}
\usepackage{pifont}
\usepackage{units}

\usepackage{tikz}
\usetikzlibrary{calc}
\usepackage{xcolor}
\usepackage{tabularx}
\usepackage{colortbl}
\usepackage{pgfplots}
\pgfplotsset{compat=newest}
\usetikzlibrary{plotmarks}
\usetikzlibrary{arrows.meta}
\usepgfplotslibrary{patchplots}
\usepackage{grffile}
\newlength\fheight 
\newlength\fwidth 
\usepgfplotslibrary{fillbetween}

%% file: commands.tex
\makeatletter
\newcommand{\gettikzxy}[3]{%
  \tikz@scan@one@point\pgfutil@firstofone#1\relax
  \edef#2{\the\pgf@x}%
  \edef#3{\the\pgf@y}%
}
\definecolor{mygray}{gray}{0.6}

%% file: acronyms.tex
\begin{acronym}[ACRONYM]
\acro{AI}{artificial intelligence}
\acro{AoA}{angle-of-arrival}
\acro{AoD}{angle-of-departure}
\acro{BS}{base station}
\acro{BP}{belief propagation}
\acro{CDF}{cumulative density function}
\acro{CFO}{carrier frequency offset}
\acro{CRB}{Cram\'er-Rao bound}
\acro{DA}{data association}
\acro{D-MIMO}{distributed multiple-input multiple-output}
\acro{DL}{downlink}
\acro{EM}{electromagnetic}
\acro{FIM}{Fisher information matrix}
\acro{GDOP}{geometric dilution of precision}
\acro{GNSS}{global navigation satellite system}
\acro{GPS}{global positioning system}
\acro{IP}{incidence point}
\acro{IQ}{in-phase and quadrature}
\acro{ISAC}{integrated sensing and communication}
\acro{ICI}{inter-carrier interference}
\acro{JCS}{Joint Communication and Sensing}
\acro{JRC}{joint radar and communication}
\acro{JRC2LS}{joint radar communication, computation, localization, and sensing}
\acro{IMU}{inertial measurement unit}
\acro{IOO}{indoor open office}
\acro{IoT}{Internet of Things}
\acro{IRN}{infrastructure reference node}
\acro{KPI}{key performance indicator}
\acro{LoS}{line-of-sight}
\acro{LS}{least-squares}
\acro{MCRB}{misspecified Cram\'er-Rao bound}
\acro{MIMO}{multiple-input multiple-output}
\acro{ML}{maximum likelihood}
\acro{mmWave}{millimeter-wave}
\acro{NLoS}{non-line-of-sight}
\acro{NR}{new radio}
\acro{OFDM}{orthogonal frequency-division multiplexing}
\acro{OTFS}{orthogonal time-frequency-space}
\acro{OEB}{orientation error bound}
\acro{PEB}{position error bound}
\acro{VEB}{velocity error bound}
\acro{PRS}{positioning reference signal}
\acro{QoS}{Quality of Service}
\acro{RAN}{radio access network}
\acro{RAT}{radio access technology}
\acro{RCS}{radar cross section}
\acro{RedCap}{reduced capacity}
\acro{RF}{radio frequency}
\acro{RIS}{reconfigurable intelligent surface}
\acro{RFS}{random finite set}
\acro{RMSE}{root mean squared error}
\acro{RTK}{real-time kinematic}
\acro{RTT}{round-trip-time}
\acro{SLAM}{simultaneous localization and mapping}
\acro{SLAT}{simultaneous localization and tracking}
\acro{SNR}{signal-to-noise ratio}
\acro{ToA}{time-of-arrival}
\acro{TDoA}{time-difference-of-arrival}
\acro{TR}{time-reversal}
\acro{TXRX}[TX/RX]{transmitter/receiver}
\acro{Tx}{transmitter}
\acro{Rx}{receiver}
\acro{UE}{user equipment}
\acro{UL}{uplink}
\acro{UWB}{ultra wideband}
\acro{XL-MIMO}{extra-large MIMO}
\end{acronym}

%% file: authors.tex
\author{Henk Wymeersch,~\IEEEmembership{Senior Member,~IEEE}, 
        Gonzalo Seco-Granados,~\IEEEmembership{Senior Member,~IEEE}
\thanks{
This work was supported by the European Commission through the H2020 project Hexa-X (Grant Agreement no.~101015956),by the ICREA Academia Program, and by the Spanish R+D project PID2020-118984GB-I00.}
\thanks{Henk Wymeersch is with the Department
of Electrical Engineering, Chalmers University of Technology, 41258 Gothenburg, Sweden (e-mail: henkw@chalmers.se). 
Gonzalo Seco-Granados is with the Department of Telecommunications and Systems Engineering, Universitat Autonoma de Barcelona, 08193 Bellaterra, Barcelona, Spain (e-mail: gonzalo.seco@uab.cat).}}

%% file: figures/FiguresPartII/letterPartII-ModelsB.tex
%
%
\definecolor{mycolor1}{rgb}{0.00000,0.44700,0.74100}%
\definecolor{mycolor2}{rgb}{0.85000,0.32500,0.09800}%
\definecolor{mycolor3}{rgb}{0.92900,0.69400,0.12500}%
\begin{tikzpicture}[scale=1\columnwidth/10cm,font=\footnotesize]

\begin{axis}[%
width=8cm,
height=3cm,
at={(1.011in,0.642in)},
scale only axis,
xmin=0.1,
xmax=10,
xlabel style={font=\color{white!15!black}},
xlabel={distance to array [m]},
ymin=-5,
ymax=0.5,
ylabel style={font=\color{white!15!black}},
ylabel={normalized response [dB]},
axis background/.style={fill=white},
legend style={at={(0.22,0.05)},anchor=south west,legend cell align=left, align=left, draw=white!15!black}
]

\addplot [color=black,line width=2.0pt]
  table[row sep=crcr]{%
0.1	0\\
0.149748743718593	0\\
0.199497487437186	0\\
0.249246231155779	0\\
0.298994974874372	0\\
0.348743718592965	0\\
0.398492462311558	0\\
0.448241206030151	0\\
0.497989949748744	0\\
0.547738693467337	0\\
0.59748743718593	0\\
0.647236180904523	0\\
0.696984924623116	0\\
0.746733668341709	0\\
0.796482412060301	0\\
0.846231155778894	0\\
0.895979899497487	0\\
0.94572864321608	0\\
0.995477386934673	0\\
1.04522613065327	0\\
1.09497487437186	0\\
1.14472361809045	0\\
1.19447236180905	0\\
1.24422110552764	0\\
1.29396984924623	0\\
1.34371859296482	0\\
1.39346733668342	0\\
1.44321608040201	0\\
1.4929648241206	0\\
1.5427135678392	0\\
1.59246231155779	0\\
1.64221105527638	0\\
1.69195979899498	0\\
1.74170854271357	0\\
1.79145728643216	0\\
1.84120603015075	0\\
1.89095477386935	0\\
1.94070351758794	0\\
1.99045226130653	0\\
2.04020100502513	0\\
2.08994974874372	0\\
2.13969849246231	0\\
2.1894472361809	0\\
2.2391959798995	0\\
2.28894472361809	0\\
2.33869346733668	0\\
2.38844221105528	0\\
2.43819095477387	0\\
2.48793969849246	0\\
2.53768844221106	0\\
2.58743718592965	0\\
2.63718592964824	0\\
2.68693467336683	0\\
2.73668341708543	0\\
2.78643216080402	0\\
2.83618090452261	0\\
2.88592964824121	0\\
2.9356783919598	0\\
2.98542713567839	0\\
3.03517587939699	0\\
3.08492462311558	0\\
3.13467336683417	0\\
3.18442211055276	0\\
3.23417085427136	0\\
3.28391959798995	0\\
3.33366834170854	0\\
3.38341708542714	0\\
3.43316582914573	0\\
3.48291457286432	0\\
3.53266331658291	0\\
3.58241206030151	0\\
3.6321608040201	0\\
3.68190954773869	0\\
3.73165829145729	0\\
3.78140703517588	0\\
3.83115577889447	0\\
3.88090452261307	0\\
3.93065326633166	0\\
3.98040201005025	0\\
4.03015075376884	0\\
4.07989949748744	0\\
4.12964824120603	0\\
4.17939698492462	0\\
4.22914572864322	0\\
4.27889447236181	0\\
4.3286432160804	0\\
4.37839195979899	0\\
4.42814070351759	0\\
4.47788944723618	0\\
4.52763819095477	0\\
4.57738693467337	0\\
4.62713567839196	0\\
4.67688442211055	0\\
4.72663316582915	0\\
4.77638190954774	0\\
4.82613065326633	0\\
4.87587939698492	0\\
4.92562814070352	0\\
4.97537688442211	0\\
5.0251256281407	0\\
5.0748743718593	0\\
5.12462311557789	0\\
5.17437185929648	0\\
5.22412060301508	0\\
5.27386934673367	0\\
5.32361809045226	0\\
5.37336683417085	0\\
5.42311557788945	0\\
5.47286432160804	0\\
5.52261306532663	0\\
5.57236180904523	0\\
5.62211055276382	0\\
5.67185929648241	0\\
5.721608040201	0\\
5.7713567839196	0\\
5.82110552763819	0\\
5.87085427135678	0\\
5.92060301507538	0\\
5.97035175879397	0\\
6.02010050251256	0\\
6.06984924623116	0\\
6.11959798994975	0\\
6.16934673366834	0\\
6.21909547738693	0\\
6.26884422110553	0\\
6.31859296482412	0\\
6.36834170854271	0\\
6.41809045226131	0\\
6.4678391959799	0\\
6.51758793969849	0\\
6.56733668341708	0\\
6.61708542713568	0\\
6.66683417085427	0\\
6.71658291457286	0\\
6.76633165829146	0\\
6.81608040201005	0\\
6.86582914572864	0\\
6.91557788944724	0\\
6.96532663316583	0\\
7.01507537688442	0\\
7.06482412060301	0\\
7.11457286432161	0\\
7.1643216080402	0\\
7.21407035175879	0\\
7.26381909547739	0\\
7.31356783919598	0\\
7.36331658291457	0\\
7.41306532663317	0\\
7.46281407035176	0\\
7.51256281407035	0\\
7.56231155778894	0\\
7.61206030150754	0\\
7.66180904522613	0\\
7.71155778894472	0\\
7.76130653266332	0\\
7.81105527638191	0\\
7.8608040201005	0\\
7.9105527638191	0\\
7.96030150753769	0\\
8.01005025125628	0\\
8.05979899497487	0\\
8.10954773869347	0\\
8.15929648241206	0\\
8.20904522613065	0\\
8.25879396984925	0\\
8.30854271356784	0\\
8.35829145728643	0\\
8.40804020100502	0\\
8.45778894472362	0\\
8.50753768844221	0\\
8.5572864321608	0\\
8.6070351758794	0\\
8.65678391959799	0\\
8.70653266331658	0\\
8.75628140703518	0\\
8.80603015075377	0\\
8.85577889447236	0\\
8.90552763819095	0\\
8.95527638190955	0\\
9.00502512562814	0\\
9.05477386934673	0\\
9.10452261306533	0\\
9.15427135678392	0\\
9.20402010050251	0\\
9.25376884422111	0\\
9.3035175879397	0\\
9.35326633165829	0\\
9.40301507537688	0\\
9.45276381909548	0\\
9.50251256281407	0\\
9.55226130653266	0\\
9.60201005025126	0\\
9.65175879396985	0\\
9.70150753768844	0\\
9.75125628140704	0\\
9.80100502512563	0\\
9.85075376884422	0\\
9.90050251256281	0\\
9.95025125628141	0\\
10	0\\
};
\addlegendentry{FF response}

\addplot [color=red,line width=2.0pt]
  table[row sep=crcr]{%
0.1	-18.6924954033254\\
0.149748743718593	-14.7228410878158\\
0.199497487437186	-14.7216303419397\\
0.249246231155779	-14.1830691248901\\
0.298994974874372	-12.2573651802091\\
0.348743718592965	-11.2680875348544\\
0.398492462311558	-10.0082922817856\\
0.448241206030151	-10.6717833802481\\
0.497989949748744	-10.72772893857\\
0.547738693467337	-10.1957074101399\\
0.59748743718593	-9.93211366032703\\
0.647236180904523	-10.0203835393128\\
0.696984924623116	-10.2397258934706\\
0.746733668341709	-10.3452672654394\\
0.796482412060301	-10.1833641896615\\
0.846231155778894	-9.74323529745094\\
0.895979899497487	-9.11551786127781\\
0.94572864321608	-8.40837730875302\\
0.995477386934673	-7.69943061618251\\
1.04522613065327	-7.0307153961637\\
1.09497487437186	-6.41978858626447\\
1.14472361809045	-5.87066002389524\\
1.19447236180905	-5.38081474426781\\
1.24422110552764	-4.94504959528783\\
1.29396984924623	-4.5573938025818\\
1.34371859296482	-4.2120018529431\\
1.39346733668342	-3.90352552698792\\
1.44321608040201	-3.62723235867106\\
1.4929648241206	-3.37900645445906\\
1.5427135678392	-3.15529929308052\\
1.59246231155779	-2.95306338035159\\
1.64221105527638	-2.76968415534654\\
1.69195979899498	-2.60291683931399\\
1.74170854271357	-2.45083065371234\\
1.79145728643216	-2.3117608102229\\
1.84120603015075	-2.18426777744949\\
1.89095477386935	-2.06710298816705\\
1.94070351758794	-1.95918007814733\\
1.99045226130653	-1.85955079604821\\
2.04020100502513	-1.76738482012709\\
2.08994974874372	-1.6819528258312\\
2.13969849246231	-1.60261225213576\\
2.1894472361809	-1.52879530712931\\
2.2391959798995	-1.45999883288537\\
2.28894472361809	-1.39577571647806\\
2.33869346733668	-1.33572758940203\\
2.38844221105528	-1.27949860323403\\
2.43819095477387	-1.22677010672126\\
2.48793969849246	-1.17725608000925\\
2.53768844221106	-1.13069920666892\\
2.58743718592965	-1.0868674845803\\
2.63718592964824	-1.0455512934229\\
2.68693467336683	-1.00656085021487\\
2.73668341708543	-0.969723995594374\\
2.78643216080402	-0.934884262802878\\
2.83618090452261	-0.901899188983458\\
2.88592964824121	-0.87063883474574\\
2.9356783919598	-0.840984483210955\\
2.98542713567839	-0.812827494129216\\
3.03517587939699	-0.786068292322942\\
3.08492462311558	-0.760615472769476\\
3.13467336683417	-0.736385007207274\\
3.18442211055276	-0.71329953931609\\
3.23417085427136	-0.691287757344879\\
3.28391959798995	-0.670283834610418\\
3.33366834170854	-0.650226929596782\\
3.38341708542714	-0.63106073850537\\
3.43316582914573	-0.612733094053361\\
3.48291457286432	-0.59519560513077\\
3.53266331658291	-0.578403332626092\\
3.58241206030151	-0.562314497323081\\
3.6321608040201	-0.546890216290367\\
3.68190954773869	-0.532094264626511\\
3.73165829145729	-0.517892859807518\\
3.78140703517588	-0.504254466214586\\
3.83115577889447	-0.49114961771172\\
3.88090452261307	-0.478550756387174\\
3.93065326633166	-0.466432085797155\\
3.98040201005025	-0.4547694372373\\
4.03015075376884	-0.443540147734266\\
4.07989949748744	-0.432722948596172\\
4.12964824120603	-0.422297863490021\\
4.17939698492462	-0.412246115122476\\
4.22914572864322	-0.402550039705713\\
4.27889447236181	-0.393193008470674\\
4.3286432160804	-0.384159355573717\\
4.37839195979899	-0.37543431180426\\
4.42814070351759	-0.367003943567574\\
4.47788944723618	-0.358855096666939\\
4.52763819095477	-0.350975344456386\\
4.57738693467337	-0.343352939982519\\
4.62713567839196	-0.33597677176421\\
4.67688442211055	-0.32883632289995\\
4.72663316582915	-0.321921633217349\\
4.77638190954774	-0.315223264209901\\
4.82613065326633	-0.308732266527201\\
4.87587939698492	-0.302440149809495\\
4.92562814070352	-0.296338854672806\\
4.97537688442211	-0.29042072667348\\
5.0251256281407	-0.284678492091786\\
5.0748743718593	-0.27910523539158\\
5.12462311557789	-0.273694378223542\\
5.17437185929648	-0.268439659853628\\
5.22412060301508	-0.263335118905355\\
5.27386934673367	-0.258375076317772\\
5.32361809045226	-0.253554119425588\\
5.37336683417085	-0.248867087079055\\
5.42311557788945	-0.244309055725409\\
5.47286432160804	-0.239875326383395\\
5.52261306532663	-0.23556141244209\\
5.57236180904523	-0.231363028230711\\
5.62211055276382	-0.227276078297233\\
5.67185929648241	-0.223296647352158\\
5.721608040201	-0.219420990826611\\
5.7713567839196	-0.215645526003601\\
5.82110552763819	-0.211966823683855\\
5.87085427135678	-0.208381600347234\\
5.92060301507538	-0.204886710779221\\
5.97035175879397	-0.201479141128495\\
6.02010050251256	-0.19815600236971\\
6.06984924623116	-0.194914524141236\\
6.11959798994975	-0.191752048936954\\
6.16934673366834	-0.188666026627181\\
6.21909547738693	-0.185654009286975\\
6.26884422110553	-0.182713646313774\\
6.31859296482412	-0.17984267981576\\
6.36834170854271	-0.177038940251949\\
6.41809045226131	-0.174300342311709\\
6.4678391959799	-0.171624881015717\\
6.51758793969849	-0.169010628026523\\
6.56733668341708	-0.16645572815575\\
6.61708542713568	-0.163958396055592\\
6.66683417085427	-0.161516913083315\\
6.71658291457286	-0.159129624330208\\
6.76633165829146	-0.156794935802506\\
6.81608040201005	-0.154511311747754\\
6.86582914572864	-0.152277272117004\\
6.91557788944724	-0.150091390155147\\
6.96532663316583	-0.147952290112197\\
7.01507537688442	-0.145858645068485\\
7.06482412060301	-0.143809174867562\\
7.11457286432161	-0.141802644150393\\
7.1643216080402	-0.13983786048512\\
7.21407035175879	-0.137913672588253\\
7.26381909547739	-0.136028968630191\\
7.31356783919598	-0.134182674622639\\
7.36331658291457	-0.132373752881889\\
7.41306532663317	-0.130601200564912\\
7.46281407035176	-0.128864048273331\\
7.51256281407035	-0.127161358722863\\
7.56231155778894	-0.125492225473962\\
7.61206030150754	-0.123855771720639\\
7.66180904522613	-0.122251149134269\\
7.71155778894472	-0.120677536760117\\
7.76130653266332	-0.119134139963184\\
7.81105527638191	-0.117620189421181\\
7.8608040201005	-0.116134940162531\\
7.9105527638191	-0.114677670646112\\
7.96030150753769	-0.113247681882043\\
8.01005025125628	-0.111844296590541\\
8.05979899497487	-0.110466858396831\\
8.10954773869347	-0.109114731061231\\
8.15929648241206	-0.107787297741648\\
8.20904522613065	-0.106483960287908\\
8.25879396984925	-0.105204138565628\\
8.30854271356784	-0.103947269808439\\
8.35829145728643	-0.102712807996953\\
8.40804020100502	-0.101500223264354\\
8.45778894472362	-0.10030900132582\\
8.50753768844221	-0.09913864293113\\
8.5572864321608	-0.0979886633403865\\
8.6070351758794	-0.0968585918201483\\
8.65678391959799	-0.0957479711605178\\
8.70653266331658	-0.0946563572110357\\
8.75628140703518	-0.0935833184355955\\
8.80603015075377	-0.09252843548414\\
8.85577889447236	-0.0914913007824239\\
8.90552763819095	-0.0904715181365191\\
8.95527638190955	-0.0894687023539391\\
9.00502512562814	-0.0884824788783982\\
9.05477386934673	-0.0875124834393528\\
9.10452261306533	-0.0865583617147024\\
9.15427135678392	-0.085619769006211\\
9.20402010050251	-0.0846963699271852\\
9.25376884422111	-0.083787838102382\\
9.3035175879397	-0.082893855878449\\
9.35326633165829	-0.0820141140461679\\
9.40301507537688	-0.0811483115717783\\
9.45276381909548	-0.0802961553389949\\
9.50251256281407	-0.0794573599001915\\
9.55226130653266	-0.0786316472370152\\
9.60201005025126	-0.0778187465292154\\
9.65175879396985	-0.0770183939321199\\
9.70150753768844	-0.076230332361824\\
9.75125628140704	-0.0754543112884887\\
9.80100502512563	-0.0746900865367393\\
9.85075376884422	-0.0739374200929097\\
9.90050251256281	-0.0731960799195324\\
9.95025125628141	-0.0724658397758926\\
10	-0.0717464790451117\\
};
\addlegendentry{NF response with FF precoder}

\addplot [color=blue, dashed,line width=2.0pt]
  table[row sep=crcr]{%
0.1	-18.6623865035707\\
0.149748743718593	-14.2387884377056\\
0.199497487437186	-15.5669949203839\\
0.249246231155779	-12.9156187992322\\
0.298994974874372	-11.0252929096194\\
0.348743718592965	-10.3030245637483\\
0.398492462311558	-11.1400786700692\\
0.448241206030151	-11.5666923292469\\
0.497989949748744	-10.4338828210111\\
0.547738693467337	-9.80168105247898\\
0.59748743718593	-9.52525705926141\\
0.647236180904523	-9.06093072994472\\
0.696984924623116	-8.23522268886764\\
0.746733668341709	-7.21049310703689\\
0.796482412060301	-6.18425846862392\\
0.846231155778894	-5.25894442054926\\
0.895979899497487	-4.46329186943198\\
0.94572864321608	-3.79244476372758\\
0.995477386934673	-3.2304718988756\\
1.04522613065327	-2.7598105777762\\
1.09497487437186	-2.36456686659033\\
1.14472361809045	-2.03134358101202\\
1.19447236180905	-1.74917218518459\\
1.24422110552764	-1.50917315905556\\
1.29396984924623	-1.30417775670969\\
1.34371859296482	-1.12838920414878\\
1.39346733668342	-0.977102224470064\\
1.44321608040201	-0.846478646690946\\
1.4929648241206	-0.733370552552153\\
1.5427135678392	-0.635181653720477\\
1.59246231155779	-0.54975868154634\\
1.64221105527638	-0.475306088066339\\
1.69195979899498	-0.410318788379937\\
1.74170854271357	-0.353528871030426\\
1.79145728643216	-0.303863151457198\\
1.84120603015075	-0.260409176317717\\
1.89095477386935	-0.222387845582834\\
1.94070351758794	-0.189131243716548\\
1.99045226130653	-0.160064593079216\\
2.04020100502513	-0.134691487094199\\
2.08994974874372	-0.112581746881622\\
2.13969849246231	-0.0933613874288165\\
2.1894472361809	-0.0767042887346633\\
2.2391959798995	-0.062325251791997\\
2.28894472361809	-0.0499741847758348\\
2.33869346733668	-0.0394312158860664\\
2.38844221105528	-0.0305025693283002\\
2.43819095477387	-0.0230170724548564\\
2.48793969849246	-0.0168231870524557\\
2.53768844221106	-0.0117864776208497\\
2.58743718592965	-0.00778744535592216\\
2.63718592964824	-0.00471966928957448\\
2.68693467336683	-0.0024882063105348\\
2.73668341708543	-0.00100821010750256\\
2.78643216080402	-0.000203735838370859\\
2.83618090452261	-6.70284950389852e-06\\
2.88592964824121	-0.000355992292531953\\
2.9356783919598	-0.00119666020634707\\
2.98542713567839	-0.0024792497029209\\
3.03517587939699	-0.00415918843891568\\
3.08492462311558	-0.0061962596687094\\
3.13467336683417	-0.00855413693674249\\
3.18442211055276	-0.0111999739405549\\
3.23417085427136	-0.0141040423322881\\
3.28391959798995	-0.0172394112660167\\
3.33366834170854	-0.0205816633759573\\
3.38341708542714	-0.0241086426122883\\
3.43316582914573	-0.0278002299908531\\
3.48291457286432	-0.0316381438485518\\
3.53266331658291	-0.035605761651544\\
3.58241206030151	-0.0396879607942306\\
3.6321608040201	-0.0438709761598871\\
3.68190954773869	-0.0481422725007365\\
3.73165829145729	-0.0524904299416587\\
3.78140703517588	-0.0569050411244282\\
3.83115577889447	-0.0613766186918097\\
3.88090452261307	-0.0658965119715505\\
3.93065326633166	-0.0704568318561012\\
3.98040201005025	-0.0750503829945937\\
4.03015075376884	-0.0796706025177311\\
4.07989949748744	-0.0843115046068513\\
4.12964824120603	-0.0889676302971583\\
4.17939698492462	-0.093634001976271\\
4.22914572864322	-0.0983060820974564\\
4.27889447236181	-0.102979735683029\\
4.3286432160804	-0.107651196237564\\
4.37839195979899	-0.112317034734958\\
4.42814070351759	-0.116974131376657\\
4.47788944723618	-0.121619649852543\\
4.52763819095477	-0.126251013864609\\
4.57738693467337	-0.130865885696003\\
4.62713567839196	-0.135462146634768\\
4.67688442211055	-0.140037879076879\\
4.72663316582915	-0.144591350154251\\
4.77638190954774	-0.149120996746679\\
4.82613065326633	-0.15362541175243\\
4.87587939698492	-0.1581033315029\\
4.92562814070352	-0.162553624220355\\
4.97537688442211	-0.166975279424261\\
5.0251256281407	-0.171367398204064\\
5.0748743718593	-0.1757291842818\\
5.12462311557789	-0.180059935796901\\
5.17437185929648	-0.18435903774981\\
5.22412060301508	-0.18862595504959\\
5.27386934673367	-0.1928602261127\\
5.32361809045226	-0.197061456968058\\
5.37336683417085	-0.201229315824885\\
5.42311557788945	-0.205363528065997\\
5.47286432160804	-0.209463871629578\\
5.52261306532663	-0.213530172751187\\
5.57236180904523	-0.21756230203121\\
5.62211055276382	-0.221560170807999\\
5.67185929648241	-0.225523727806149\\
5.721608040201	-0.229452956041977\\
5.7713567839196	-0.233347869964179\\
5.82110552763819	-0.237208512810514\\
5.87085427135678	-0.241034954165727\\
5.92060301507538	-0.244827287702277\\
5.97035175879397	-0.24858562909236\\
6.02010050251256	-0.252310114075007\\
6.06984924623116	-0.256000896670332\\
6.11959798994975	-0.259658147525535\\
6.16934673366834	-0.263282052384933\\
6.21909547738693	-0.266872810675074\\
6.26884422110553	-0.270430634194832\\
6.31859296482412	-0.273955745902473\\
6.36834170854271	-0.27744837879523\\
6.41809045226131	-0.280908774870133\\
6.4678391959799	-0.284337184164137\\
6.51758793969849	-0.287733863865426\\
6.56733668341708	-0.291099077490996\\
6.61708542713568	-0.29443309412618\\
6.66683417085427	-0.297736187721965\\
6.71658291457286	-0.301008636443573\\
6.76633165829146	-0.304250722070268\\
6.81608040201005	-0.307462729439503\\
6.86582914572864	-0.310644945933759\\
6.91557788944724	-0.313797661006788\\
6.96532663316583	-0.31692116574622\\
7.01507537688442	-0.320015752470121\\
7.06482412060301	-0.323081714354749\\
7.11457286432161	-0.326119345091864\\
7.1643216080402	-0.329128938573503\\
7.21407035175879	-0.33211078860067\\
7.26381909547739	-0.335065188617275\\
7.31356783919598	-0.3379924314643\\
7.36331658291457	-0.340892809155115\\
7.41306532663317	-0.343766612668975\\
7.46281407035176	-0.34661413176289\\
7.51256281407035	-0.349435654798807\\
7.56231155778894	-0.352231468586226\\
7.61206030150754	-0.355001858238888\\
7.66180904522613	-0.357747107044878\\
7.71155778894472	-0.360467496348035\\
7.76130653266332	-0.363163305441308\\
7.81105527638191	-0.365834811470387\\
7.8608040201005	-0.36848228934651\\
7.9105527638191	-0.371106011669587\\
7.96030150753769	-0.373706248658337\\
8.01005025125628	-0.376283268088683\\
8.05979899497487	-0.37883733524\\
8.10954773869347	-0.38136871284708\\
8.15929648241206	-0.383877661059227\\
8.20904522613065	-0.3863644374042\\
8.25879396984925	-0.388829296758053\\
8.30854271356784	-0.391272491319849\\
8.35829145728643	-0.393694270591734\\
8.40804020100502	-0.396094881361311\\
8.45778894472362	-0.398474567689487\\
8.50753768844221	-0.400833570902294\\
8.5572864321608	-0.403172129584388\\
8.6070351758794	-0.405490479577402\\
8.65678391959799	-0.407788853979726\\
8.70653266331658	-0.410067483150137\\
8.75628140703518	-0.412326594712651\\
8.80603015075377	-0.414566413565284\\
8.85577889447236	-0.416787161888558\\
8.90552763819095	-0.418989059158451\\
8.95527638190955	-0.421172322158556\\
9.00502512562814	-0.423337164995949\\
9.05477386934673	-0.425483799117183\\
9.10452261306533	-0.427612433325755\\
9.15427135678392	-0.429723273801409\\
9.20402010050251	-0.431816524120335\\
9.25376884422111	-0.433892385275426\\
9.3035175879397	-0.435951055699059\\
9.35326633165829	-0.437992731284175\\
9.40301507537688	-0.440017605409006\\
9.45276381909548	-0.442025868960595\\
9.50251256281407	-0.444017710359476\\
9.55226130653266	-0.445993315584099\\
9.60201005025126	-0.447952868196718\\
9.65175879396985	-0.449896549369034\\
9.70150753768844	-0.451824537908662\\
9.75125628140704	-0.45373701028505\\
9.80100502512563	-0.455634140655882\\
9.85075376884422	-0.457516100894584\\
9.90050251256281	-0.459383060616445\\
9.95025125628141	-0.46123518720617\\
10	-0.46307264584464\\
};
\addlegendentry{NF response with NF precoder}
\addplot[color=black,line width=0.5pt,dotted]
table[row sep=crcr]{%
2.8 -5\\
2.8 0.5\\
};
\end{axis}
\end{tikzpicture}%

%% file: figures/FiguresPartII/letterPartII-AF.tex
%
%
\definecolor{mycolor1}{rgb}{0.00000,0.44700,0.74100}%
\definecolor{mycolor2}{rgb}{0.85000,0.32500,0.09800}%
\begin{tikzpicture}[scale=1\columnwidth/10cm,font=\footnotesize]

\begin{axis}[%
width=8.3cm,
height=3cm,
at={(1.011in,0.642in)},
scale only axis,
xmin=-10,
xmax=30,
xlabel style={font=\color{white!15!black}},
xlabel={distance [m]},
ymin=-20,
ymax=32,
ylabel style={font=\color{white!15!black}},
ylabel={magnitude [dB]},
legend style={at={(0.22,0.65)},anchor=south west,legend cell align=left, align=left, draw=white!15!black},
axis background/.style={fill=white}
]
\addplot [color=mycolor1,style=dashed,line width=2.0pt]
  table[row sep=crcr]{%
-11.4772727272727	0.855746280980011\\
-11.3636363636363	0.57354221916467\\
-11.25	0.0533471736314425\\
-11.1363636363635	-0.738555381165447\\
-11.022727272727	-1.86119486568692\\
-10.9090909090908	-3.41945398868828\\
-10.7954545454545	-5.61556856630984\\
-10.6818181818182	-8.90927332205693\\
-10.5681818181818	-14.7744262305869\\
-10.4545454545453	-289.570285907962\\
-10.340909090909	-14.6779366081905\\
-10.2272727272725	-8.71628811931232\\
-10.1136363636363	-5.32607586510642\\
-9.99999999999977	-3.03343590634013\\
-9.88636363636351	-1.37862754830335\\
-9.77272727272725	-0.159408994802\\
-9.65909090909076	0.729108454049492\\
-9.5454545454545	1.34596022315696\\
-9.43181818181802	1.72486885807576\\
-9.31818181818153	1.88340403030892\\
-9.20454545454527	1.8270368992064\\
-9.09090909090901	1.55030337791255\\
-8.97727272727275	1.03564087035613\\
-8.86363636363626	0.249333912291347\\
-8.74999999999977	-0.867645831053772\\
-8.63636363636351	-2.42017995765074\\
-8.52272727272702	-4.61050314703153\\
-8.40909090909076	-7.89834895951274\\
-8.2954545454545	-13.7575741792001\\
-8.18181818181802	-293.402044940666\\
-8.06818181818176	-13.649018038477\\
-7.95454545454527	-7.68122819475835\\
-7.84090909090901	-4.28480078765219\\
-7.72727272727252	-1.98587053780696\\
-7.61363636363626	-0.324695377708826\\
-7.5	0.900967895328623\\
-7.38636363636351	1.79600942241319\\
-7.27272727272725	2.41946610548253\\
-7.15909090909076	2.80506200405426\\
-7.04545454545428	2.97036834168675\\
-6.93181818181802	2.920857869086\\
-6.81818181818176	2.65106813130638\\
-6.7045454545455	2.14343820604259\\
-6.59090909090901	1.36425434602307\\
-6.47727272727252	0.254489978063802\\
-6.36363636363626	-1.29073468814474\\
-6.24999999999977	-3.47365247687993\\
-6.13636363636351	-6.75399504403247\\
-6.02272727272725	-12.6056172184683\\
-5.90909090909076	-295.734587283724\\
-5.7954545454545	-12.4815475021244\\
-5.68181818181802	-6.50584294775997\\
-5.56818181818176	-3.10139266575308\\
-5.45454545454527	-0.794329140373015\\
-5.34090909090901	0.875091994451\\
-5.22727272727275	2.10911630431955\\
-5.11363636363626	3.01263635988355\\
-5	3.644691561622\\
-4.88636363636351	4.0390085390629\\
-4.77272727272702	4.21316115982237\\
-4.65909090909076	4.17262489540207\\
-4.5454545454545	3.91194009119923\\
-4.43181818181824	3.4135487077322\\
-4.31818181818176	2.64373996609864\\
-4.20454545454527	1.54349035023185\\
-4.09090909090901	0.00792321903540717\\
-3.97727272727252	-2.16519100695786\\
-3.86363636363626	-5.43558063956986\\
-3.74999999999977	-11.2770970605069\\
-3.63636363636351	-288.908366810976\\
-3.52272727272725	-11.1323430519412\\
-3.40909090909076	-5.1460525120232\\
-3.2954545454545	-1.7308485228322\\
-3.18181818181802	0.587140458089107\\
-3.06818181818153	2.267662953754\\
-2.95454545454527	3.51296882223913\\
-2.84090909090901	4.42795506778632\\
-2.72727272727275	5.0716656714402\\
-2.61363636363626	5.47783199642408\\
-2.49999999999977	5.66403280364668\\
-2.38636363636351	5.63574862429856\\
-2.27272727272702	5.38752503701233\\
-2.15909090909076	4.90180941661151\\
-2.0454545454545	4.14489658748731\\
-1.93181818181802	3.05776883419434\\
-1.81818181818176	1.53555552237097\\
-1.70454545454527	-0.623966705341755\\
-1.59090909090901	-3.88051971351755\\
-1.47727272727252	-9.70794820117208\\
-1.36363636363626	-285.067550509844\\
-1.25	-9.53423642566016\\
-1.13636363636351	-3.53306140850935\\
-1.02272727272725	-0.102692321133859\\
-0.909090909090764	2.23075041479384\\
-0.795454545454277	3.92702366886375\\
-0.681818181818016	5.18838597164478\\
-0.568181818181756	6.11974334190089\\
-0.454545454545496	6.78014913431133\\
-0.340909090909008	7.20334446296669\\
-0.227272727272521	7.40691823610778\\
-0.11363636363626	7.39636154914666\\
2.27373675443232e-13	7.16623098353167\\
0.113636363636488	6.69898537857592\\
0.227272727272748	5.96093150939196\\
0.340909090909236	4.89306412363487\\
0.454545454545496	3.39052559028265\\
0.568181818181984	1.2511061252592\\
0.681818181818244	-1.9849019610903\\
0.795454545454731	-7.79132855738558\\
0.909090909090992	-283.089564297644\\
1.02272727272725	-7.57417945212706\\
1.13636363636374	-1.55053586023726\\
1.25	1.90282512981319\\
1.36363636363649	4.25980167988035\\
1.47727272727298	5.98017013755597\\
1.59090909090924	7.26620946060599\\
1.7045454545455	8.22284710299394\\
1.81818181818176	8.90915892585642\\
1.93181818181824	9.35890968910151\\
2.04545454545473	9.58971315953793\\
2.15909090909099	9.60708658306215\\
2.27272727272748	9.40561406900546\\
2.38636363636374	8.96778345435898\\
2.50000000000023	8.2599320816844\\
2.61363636363649	7.22308694415267\\
2.72727272727275	5.75242445176146\\
2.84090909090924	3.64577078445705\\
2.9545454545455	0.443456278195483\\
3.06818181818198	-5.32830871762358\\
3.18181818181847	-284.231463760273\\
3.29545454545473	-5.03875933510347\\
3.40909090909099	1.02271602719696\\
3.52272727272725	4.51506340552139\\
3.63636363636374	6.91223505190494\\
3.75000000000023	8.6740648628617\\
3.86363636363649	10.0028930869459\\
3.97727272727298	11.0037125223896\\
4.09090909090924	11.73566877161\\
4.2045454545455	12.2326011121698\\
4.31818181818198	12.512203024569\\
4.43181818181824	12.5800771295234\\
4.54545454545473	12.4308990887245\\
4.65909090909099	12.0472550435862\\
4.77272727272748	11.3955880336339\\
4.88636363636374	10.4170388570644\\
5	9.00690663809439\\
5.11363636363649	6.96315007913279\\
5.22727272727275	3.82624285821475\\
5.34090909090924	-1.87744937297484\\
5.45454545454572	-277.267629861187\\
5.56818181818198	-1.44306910233216\\
5.68181818181824	4.69554728739238\\
5.7954545454545	8.2684705397182\\
5.90909090909099	10.7498912595836\\
6.02272727272748	12.5999046294545\\
6.13636363636374	14.0211379870863\\
6.25000000000023	15.1189003421859\\
6.36363636363649	15.952686484058\\
6.47727272727275	16.5567223444047\\
6.59090909090924	16.9491307846687\\
6.7045454545455	17.1359927206978\\
6.81818181818198	17.1125183387853\\
6.93181818181824	16.8618932293242\\
7.04545454545473	16.3512351988619\\
7.15909090909099	15.5224476309526\\
7.27272727272725	14.271695186217\\
7.38636363636374	12.3979235038121\\
7.5	9.4427372639763\\
7.61363636363649	3.93380526179708\\
7.72727272727298	-277.109544005024\\
7.84090909090924	4.80311677581179\\
7.9545454545455	11.1857360551399\\
8.06818181818176	15.0234947117769\\
8.18181818181824	17.7935109209016\\
8.29545454545473	19.9594108152434\\
8.40909090909099	21.7281272750696\\
8.52272727272748	23.2102849351157\\
8.63636363636374	24.472035151818\\
8.75000000000023	25.5560777210453\\
8.86363636363649	26.4915322629868\\
8.97727272727275	27.2990800546202\\
9.09090909090924	27.9938579713254\\
9.2045454545455	28.587187638558\\
9.31818181818198	29.087658257749\\
9.43181818181847	29.5018300056209\\
9.54545454545473	29.8347036626418\\
9.65909090909099	30.0900397793267\\
9.77272727272725	30.2705767975007\\
9.88636363636374	30.378178167639\\
10.0000000000002	30.4139268515823\\
10.1136363636365	30.378178167639\\
10.227272727273	30.2705767975007\\
10.3409090909092	30.0900397793267\\
10.4545454545455	29.8347036626418\\
10.568181818182	29.5018300056209\\
10.6818181818182	29.087658257749\\
10.7954545454547	28.5871876385579\\
10.909090909091	27.9938579713254\\
11.0227272727275	27.2990800546202\\
11.1363636363637	26.4915322629868\\
11.25	25.5560777210453\\
11.3636363636365	24.472035151818\\
11.4772727272727	23.2102849351157\\
11.5909090909092	21.7281272750696\\
11.7045454545457	19.9594108152434\\
11.818181818182	17.7935109209016\\
11.9318181818182	15.0234947117768\\
12.0454545454545	11.1857360551399\\
12.159090909091	4.80311677581161\\
12.2727272727275	-273.357646389331\\
12.3863636363637	3.93380526179714\\
12.5000000000002	9.44273726397638\\
12.6136363636365	12.3979235038122\\
12.7272727272727	14.271695186217\\
12.8409090909092	15.5224476309527\\
12.9545454545455	16.3512351988619\\
13.068181818182	16.8618932293242\\
13.1818181818182	17.1125183387853\\
13.2954545454547	17.1359927206978\\
13.409090909091	16.9491307846687\\
13.5227272727273	16.5567223444047\\
13.6363636363637	15.952686484058\\
13.75	15.1189003421859\\
13.8636363636365	14.0211379870863\\
13.977272727273	12.5999046294545\\
14.0909090909092	10.7498912595836\\
14.2045454545455	8.26847053971811\\
14.3181818181818	4.6955472873923\\
14.4318181818182	-1.44306910233227\\
14.5454545454547	-281.379889432666\\
14.659090909091	-1.87744937297477\\
14.7727272727275	3.82624285821482\\
14.8863636363637	6.96315007913283\\
15.0000000000002	9.00690663809441\\
15.1136363636365	10.4170388570644\\
15.2272727272727	11.3955880336339\\
15.3409090909092	12.0472550435862\\
15.4545454545455	12.4308990887245\\
15.568181818182	12.5800771295234\\
15.6818181818185	12.512203024569\\
15.7954545454547	12.2326011121698\\
15.909090909091	11.73566877161\\
16.0227272727273	11.0037125223895\\
16.1363636363637	10.0028930869459\\
16.2500000000002	8.67406486286163\\
16.3636363636365	6.9122350519049\\
16.477272727273	4.51506340552137\\
16.5909090909092	1.0227160271969\\
16.7045454545455	-5.03875933510363\\
16.818181818182	-286.508403982807\\
16.9318181818182	-5.32830871762346\\
17.0454545454547	0.44345627819551\\
17.159090909091	3.64577078445711\\
17.2727272727275	5.75242445176146\\
17.3863636363637	7.22308694415269\\
17.5	8.25993208168442\\
17.6136363636365	8.96778345435897\\
17.7272727272727	9.40561406900547\\
17.8409090909092	9.60708658306214\\
17.9545454545457	9.58971315953793\\
18.068181818182	9.3589096891015\\
18.1818181818182	8.90915892585641\\
18.2954545454545	8.22284710299391\\
18.409090909091	7.26620946060599\\
18.5227272727275	5.98017013755592\\
18.6363636363637	4.2598016798803\\
18.7500000000002	1.90282512981312\\
18.8636363636365	-1.55053586023734\\
18.9772727272727	-7.57417945212713\\
19.0909090909092	-299.206972271901\\
19.2045454545455	-7.79132855738535\\
19.318181818182	-1.9849019610902\\
19.4318181818182	1.2511061252592\\
19.5454545454547	3.3905255902827\\
19.659090909091	4.8930641236349\\
19.7727272727273	5.96093150939196\\
19.8863636363637	6.69898537857593\\
20	7.16623098353168\\
20.1136363636365	7.39636154914667\\
20.227272727273	7.40691823610778\\
20.3409090909092	7.20334446296667\\
20.4545454545455	6.78014913431131\\
20.5681818181818	6.11974334190087\\
20.6818181818182	5.18838597164478\\
20.7954545454547	3.92702366886372\\
20.909090909091	2.23075041479384\\
21.0227272727275	-0.102692321133917\\
21.1363636363637	-3.53306140850942\\
21.2500000000002	-9.53423642566026\\
21.3636363636365	-293.638422172954\\
21.4772727272727	-9.70794820117189\\
21.5909090909092	-3.88051971351751\\
21.7045454545455	-0.623966705341708\\
21.818181818182	1.53555552237097\\
21.9318181818185	3.05776883419437\\
22.0454545454547	4.14489658748732\\
22.159090909091	4.90180941661151\\
22.2727272727273	5.38752503701235\\
22.3863636363637	5.63574862429857\\
22.5000000000002	5.66403280364672\\
22.6136363636365	5.47783199642407\\
22.727272727273	5.07166567144019\\
22.8409090909092	4.42795506778629\\
22.9545454545455	3.5129688222391\\
23.068181818182	2.26766295375397\\
23.1818181818182	0.587140458089055\\
23.2954545454547	-1.73084852283224\\
23.409090909091	-5.14605251202327\\
23.5227272727275	-11.1323430519413\\
23.6363636363637	-297.65844483066\\
23.75	-11.2770970605068\\
23.8636363636365	-5.43558063956978\\
23.9772727272727	-2.16519100695782\\
24.0909090909092	0.0079232190354399\\
24.2045454545457	1.54349035023185\\
24.318181818182	2.64373996609866\\
24.4318181818182	3.41354870773222\\
24.5454545454545	3.91194009119924\\
24.659090909091	4.17262489540207\\
24.7727272727275	4.21316115982238\\
24.8863636363637	4.03900853906289\\
25.0000000000002	3.644691561622\\
25.1136363636365	3.01263635988351\\
25.2272727272727	2.10911630431955\\
25.3409090909092	0.875091994450964\\
25.4545454545455	-0.794329140373062\\
25.568181818182	-3.10139266575307\\
25.6818181818182	-6.50584294776\\
25.7954545454547	-12.4815475021245\\
25.909090909091	-286.590798277672\\
26.0227272727273	-12.6056172184682\\
26.1363636363637	-6.75399504403243\\
26.25	-3.47365247687994\\
26.3636363636365	-1.29073468814472\\
26.477272727273	0.254489978063803\\
26.5909090909092	1.36425434602309\\
26.7045454545455	2.1434382060426\\
26.8181818181818	2.65106813130638\\
26.9318181818182	2.92085786908598\\
27.0454545454547	2.97036834168673\\
27.159090909091	2.80506200405423\\
27.2727272727275	2.41946610548252\\
27.3863636363637	1.79600942241317\\
27.5000000000002	0.90096789532862\\
27.6136363636365	-0.324695377708882\\
27.7272727272727	-1.985870537807\\
27.8409090909092	-4.28480078765222\\
27.9545454545455	-7.68122819475845\\
28.068181818182	-13.649018038477\\
28.1818181818185	-289.077276771536\\
28.2954545454547	-13.7575741792\\
28.409090909091	-7.89834895951267\\
28.5227272727273	-4.61050314703147\\
28.6363636363637	-2.42017995765069\\
28.7500000000002	-0.867645831053768\\
28.8636363636365	0.249333912291331\\
28.977272727273	1.03564087035615\\
29.0909090909092	1.55030337791255\\
29.2045454545455	1.82703689920639\\
29.318181818182	1.88340403030891\\
29.4318181818182	1.72486885807575\\
29.5454545454547	1.34596022315693\\
29.659090909091	0.729108454049486\\
29.7727272727275	-0.159408994802012\\
29.8863636363637	-1.37862754830337\\
30	-3.03343590634018\\
30.1136363636365	-5.32607586510648\\
30.2272727272727	-8.71628811931241\\
30.3409090909092	-14.6779366081906\\
30.4545454545457	-296.026074062508\\
30.568181818182	-14.7744262305867\\
30.6818181818182	-8.90927332205686\\
30.7954545454545	-5.61556856630975\\
30.909090909091	-3.41945398868826\\
31.0227272727275	-1.86119486568691\\
31.1363636363637	-0.738555381165427\\
31.2500000000002	0.0533471736314415\\
31.3636363636365	0.573542219164672\\
31.4772727272727	0.855746280980006\\
31.5909090909092	0.917522992741127\\
31.7045454545455	0.764337455847644\\
31.818181818182	0.390719503082067\\
31.9318181818182	-0.220899566721427\\
32.0454545454547	-1.10424134998058\\
32.159090909091	-2.31834034308681\\
32.2727272727273	-3.96808433747681\\
32.3863636363637	-6.25571424042167\\
32.5	-9.6409698767122\\
32.6136363636365	-15.5977143341596\\
32.727272727273	-291.944821100972\\
32.8409090909092	-15.6845503370317\\
32.9545454545457	-9.81464622572694\\
33.068181818182	-6.51623962342295\\
33.1818181818182	-4.31547178948955\\
33.2954545454547	-2.75260725023047\\
33.409090909091	-1.62540945469172\\
33.5227272727275	-0.828994974289393\\
33.6363636363637	-0.304333683338428\\
33.7500000000002	-0.0177083653047651\\
33.8636363636365	0.0484452905855638\\
33.9772727272727	-0.100406950688688\\
};
\addplot [color=mycolor2,line width=2.0pt]
  table[row sep=crcr]{%
-11.4772727272727	13.3843842985718\\
-11.3636363636363	13.5478504875495\\
-11.25	13.500682317966\\
-11.1363636363635	13.2385422158496\\
-11.022727272727	12.7457002160009\\
-10.9090909090908	11.9915219460657\\
-10.7954545454545	10.9225049601356\\
-10.6818181818182	9.4446990138402\\
-10.5681818181818	7.38092156723677\\
-10.4545454545453	4.34435981023722\\
-10.340909090909	-0.798013572997445\\
-10.2272727272725	-16.7058055598313\\
-10.1136363636363	-3.99813909326765\\
-9.99999999999977	2.92729254509049\\
-9.88636363636351	6.62547159758948\\
-9.77272727272725	9.07654646369559\\
-9.65909090909076	10.8334080484633\\
-9.5454545454545	12.1286173056846\\
-9.43181818181802	13.0801704414987\\
-9.31818181818153	13.7540131661285\\
-9.20454545454527	14.1882292974618\\
-9.09090909090901	14.4037258360179\\
-8.97727272727275	14.4091125397213\\
-8.86363636363626	14.2025107138338\\
-8.74999999999977	13.7711644966907\\
-8.63636363636351	13.0886736725724\\
-8.52272727272702	12.1084540789577\\
-8.40909090909076	10.7494517032786\\
-8.2954545454545	8.86253696487397\\
-8.18181818181802	6.13707308002727\\
-8.06818181818176	1.74886795257391\\
-7.95454545454527	-8.38144770194792\\
-7.84090909090901	-6.38774317356965\\
-7.72727272727252	2.52527316226289\\
-7.61363636363626	6.73989729788372\\
-7.5	9.43996872706655\\
-7.38636363636351	11.3508240553516\\
-7.27272727272725	12.756147984829\\
-7.15909090909076	13.794686938913\\
-7.04545454545428	14.5426229187681\\
-6.93181818181802	15.0439235676295\\
-6.81818181818176	15.3234218407607\\
-6.7045454545455	15.3928277627812\\
-6.59090909090901	15.2532252093708\\
-6.47727272727252	14.8952885496887\\
-6.36363636363626	14.2973189945604\\
-6.24999999999977	13.4201521953399\\
-6.13636363636351	12.1960614414409\\
-6.02272727272725	10.5035542045347\\
-5.90909090909076	8.10169841338259\\
-5.7954545454545	4.41063213944918\\
-5.68181818181802	-2.71358351540002\\
-5.56818181818176	-12.8896599044133\\
-5.45454545454527	1.51889313037989\\
-5.34090909090901	6.60205924154729\\
-5.22727272727275	9.67488880700504\\
-5.11363636363626	11.8032631855452\\
-5	13.35848841692\\
-4.88636363636351	14.5123412790102\\
-4.77272727272702	15.3565000520001\\
-4.65909090909076	15.9434669997372\\
-4.5454545454545	16.3035198695487\\
-4.43181818181824	16.4524484742475\\
-4.31818181818176	16.3950126906708\\
-4.20454545454527	16.1259051617722\\
-4.09090909090901	15.6286636159955\\
-3.97727272727252	14.8720281073956\\
-3.86363636363626	13.8018457050268\\
-3.74999999999977	12.3232754917953\\
-3.63636363636351	10.2574640893986\\
-3.52272727272725	7.21319451430435\\
-3.40909090909076	2.03851611725471\\
-3.2954545454545	-14.3594977092596\\
-3.18181818181802	-0.914226349180115\\
-3.06818181818153	5.95322493622528\\
-2.95454545454527	9.64675359386411\\
-2.84090909090901	12.1061283286938\\
-2.72727272727275	13.8775235735146\\
-2.61363636363626	15.1915650100126\\
-2.49999999999977	16.1656454966753\\
-2.38636363636351	16.8656583759395\\
-2.27272727272702	17.3299668464552\\
-2.15909090909076	17.5800492292918\\
-2.0454545454545	17.6254322612404\\
-1.93181818181802	17.4656450630911\\
-1.81818181818176	17.0901180931633\\
-1.70454545454527	16.4759814250997\\
-1.59090909090901	15.5827081372841\\
-1.47727272727252	14.340568110588\\
-1.36363636363626	12.6243638952642\\
-1.25	10.1844645671491\\
-1.13636363636351	6.41276715491086\\
-1.02272727272725	-1.01459965943983\\
-0.909090909090764	-9.10516576297968\\
-0.795454545454277	4.08329688746914\\
-0.681818181818016	9.03630504487097\\
-0.568181818181756	12.0667504463812\\
-0.454545454545496	14.1801237151219\\
-0.340909090909008	15.7334247536094\\
-0.227272727272521	16.8934377128605\\
-0.11363636363626	17.7498065555731\\
2.27373675443232e-13	18.3542433118692\\
0.113636363636488	18.7369103722316\\
0.227272727272748	18.9139725320404\\
0.340909090909236	18.8910514244987\\
0.454545454545496	18.6643261250491\\
0.568181818181984	18.2197736370921\\
0.681818181818244	17.5302113969781\\
0.795454545454731	16.5486898760715\\
0.909090909090992	15.1942759215355\\
1.02272727272725	13.3188090443327\\
1.13636363636374	10.6148641662392\\
1.25	6.27101519083982\\
1.36363636363649	-3.66486189179775\\
1.47727272727298	-2.12643760263582\\
1.59090909090924	6.94702993456474\\
1.7045454545455	11.2178495882814\\
1.81818181818176	13.9616080964908\\
1.93181818181824	15.9150490630396\\
2.04545454545473	17.3654286512311\\
2.15909090909099	18.4533418535405\\
2.27272727272748	19.2563965761513\\
2.38636363636374	19.8200160845362\\
2.50000000000023	20.1707816519672\\
2.61363636363649	20.3227032227413\\
2.72727272727275	20.2800756566422\\
2.84090909090924	20.0382859935768\\
2.9545454545455	19.582919844345\\
3.06818181818198	18.886779782879\\
3.18181818181847	17.9033784171656\\
3.29545454545473	16.553062496843\\
3.40909090909099	14.6907790152754\\
3.52272727272725	12.017564931176\\
3.63636363636374	7.75375854662635\\
3.75000000000023	-1.75267822131566\\
3.86363636363649	-1.42081991885115\\
3.97727272727298	8.03189819820573\\
4.09090909090924	12.385972883072\\
4.2045454545455	15.1707440131249\\
4.31818181818198	17.1525589793351\\
4.43181818181824	18.6270491911236\\
4.54545454545473	19.7382916198684\\
4.65909090909099	20.5658160742003\\
4.77272727272748	21.1564266781177\\
4.88636363636374	21.5379673599779\\
5	21.7258428507113\\
5.11363636363649	21.7261026843118\\
5.22727272727275	21.5365458712208\\
5.34090909090924	21.1463084431922\\
5.45454545454572	20.5337646407304\\
5.56818181818198	19.6618119245559\\
5.68181818181824	18.468058303608\\
5.7954545454545	16.8432258377867\\
5.90909090909099	14.5769301707862\\
6.02272727272748	11.1869798943317\\
6.13636363636374	5.0877921826408\\
6.25000000000023	-20.6511940366074\\
6.36363636363649	6.03041097815539\\
6.47727272727275	11.8025109157939\\
6.59090909090924	15.1483777452977\\
6.7045454545455	17.4419516152857\\
6.81818181818198	19.1234199495482\\
6.93181818181824	20.3894288969829\\
7.04545454545473	21.3436396562073\\
7.15909090909099	22.0459361397601\\
7.27272727272725	22.5323787673382\\
7.38636363636374	22.8243995836729\\
7.5	22.9332953282899\\
7.61363636363649	22.8623075008125\\
7.72727272727298	22.6071777608192\\
7.84090909090924	22.1553956321548\\
7.9545454545455	21.4838419223941\\
8.06818181818176	20.5537997889454\\
8.18181818181824	19.3007061158338\\
8.29545454545473	17.6115833911018\\
8.40909090909099	15.2679650203353\\
8.52272727272748	11.7635561815074\\
8.63636363636374	5.38573099884934\\
8.75000000000023	-13.7944308787457\\
8.86363636363649	7.13276106811567\\
8.97727272727275	12.6790027895423\\
9.09090909090924	15.9270320600602\\
9.2045454545455	18.1574687757705\\
9.31818181818198	19.7907635024025\\
9.43181818181847	21.0168555004495\\
9.54545454545473	21.9365471385502\\
9.65909090909099	22.6083611733256\\
9.77272727272725	23.0677095019972\\
9.88636363636374	23.3357871572316\\
10.0000000000002	23.4239521074884\\
10.1136363636365	23.3357871572316\\
10.227272727273	23.0677095019972\\
10.3409090909092	22.6083611733255\\
10.4545454545455	21.9365471385502\\
10.568181818182	21.0168555004495\\
10.6818181818182	19.7907635024025\\
10.7954545454547	18.1574687757705\\
10.909090909091	15.9270320600602\\
11.0227272727275	12.6790027895422\\
11.1363636363637	7.13276106811556\\
11.25	-13.794430878747\\
11.3636363636365	5.38573099884945\\
11.4772727272727	11.7635561815075\\
11.5909090909092	15.2679650203354\\
11.7045454545457	17.6115833911019\\
11.818181818182	19.3007061158338\\
11.9318181818182	20.5537997889455\\
12.0454545454545	21.4838419223941\\
12.159090909091	22.1553956321548\\
12.2727272727275	22.6071777608192\\
12.3863636363637	22.8623075008125\\
12.5000000000002	22.9332953282899\\
12.6136363636365	22.824399583673\\
12.7272727272727	22.5323787673382\\
12.8409090909092	22.0459361397601\\
12.9545454545455	21.3436396562073\\
13.068181818182	20.3894288969828\\
13.1818181818182	19.1234199495481\\
13.2954545454547	17.4419516152857\\
13.409090909091	15.1483777452977\\
13.5227272727273	11.8025109157939\\
13.6363636363637	6.03041097815528\\
13.75	-20.6511940366102\\
13.8636363636365	5.08779218264097\\
13.977272727273	11.1869798943317\\
14.0909090909092	14.5769301707862\\
14.2045454545455	16.8432258377867\\
14.3181818181818	18.468058303608\\
14.4318181818182	19.6618119245559\\
14.5454545454547	20.5337646407304\\
14.659090909091	21.1463084431922\\
14.7727272727275	21.5365458712208\\
14.8863636363637	21.7261026843119\\
15.0000000000002	21.7258428507113\\
15.1136363636365	21.5379673599779\\
15.2272727272727	21.1564266781177\\
15.3409090909092	20.5658160742003\\
15.4545454545455	19.7382916198684\\
15.568181818182	18.6270491911235\\
15.6818181818185	17.1525589793351\\
15.7954545454547	15.1707440131249\\
15.909090909091	12.3859728830719\\
16.0227272727273	8.03189819820563\\
16.1363636363637	-1.42081991885144\\
16.2500000000002	-1.75267822131531\\
16.3636363636365	7.7537585466265\\
16.477272727273	12.017564931176\\
16.5909090909092	14.6907790152754\\
16.7045454545455	16.553062496843\\
16.818181818182	17.9033784171656\\
16.9318181818182	18.886779782879\\
17.0454545454547	19.582919844345\\
17.159090909091	20.0382859935768\\
17.2727272727275	20.2800756566422\\
17.3863636363637	20.3227032227413\\
17.5	20.1707816519672\\
17.6136363636365	19.8200160845362\\
17.7272727272727	19.2563965761513\\
17.8409090909092	18.4533418535404\\
17.9545454545457	17.3654286512311\\
18.068181818182	15.9150490630396\\
18.1818181818182	13.9616080964908\\
18.2954545454545	11.2178495882814\\
18.409090909091	6.94702993456466\\
18.5227272727275	-2.12643760263624\\
18.6363636363637	-3.66486189179742\\
18.7500000000002	6.27101519083994\\
18.8636363636365	10.6148641662393\\
18.9772727272727	13.3188090443328\\
19.0909090909092	15.1942759215355\\
19.2045454545455	16.5486898760716\\
19.318181818182	17.5302113969781\\
19.4318181818182	18.2197736370921\\
19.5454545454547	18.6643261250491\\
19.659090909091	18.8910514244986\\
19.7727272727273	18.9139725320404\\
19.8863636363637	18.7369103722316\\
20	18.3542433118692\\
20.1136363636365	17.7498065555731\\
20.227272727273	16.8934377128605\\
20.3409090909092	15.7334247536093\\
20.4545454545455	14.1801237151218\\
20.5681818181818	12.0667504463811\\
20.6818181818182	9.03630504487093\\
20.7954545454547	4.08329688746898\\
20.909090909091	-9.10516576298021\\
21.0227272727275	-1.01459965943966\\
21.1363636363637	6.41276715491098\\
21.2500000000002	10.1844645671491\\
21.3636363636365	12.6243638952643\\
21.4772727272727	14.3405681105881\\
21.5909090909092	15.5827081372841\\
21.7045454545455	16.4759814250997\\
21.818181818182	17.0901180931633\\
21.9318181818185	17.4656450630911\\
22.0454545454547	17.6254322612404\\
22.159090909091	17.5800492292918\\
22.2727272727273	17.3299668464552\\
22.3863636363637	16.8656583759395\\
22.5000000000002	16.1656454966753\\
22.6136363636365	15.1915650100125\\
22.727272727273	13.8775235735146\\
22.8409090909092	12.1061283286938\\
22.9545454545455	9.64675359386405\\
23.068181818182	5.95322493622518\\
23.1818181818182	-0.914226349180333\\
23.2954545454547	-14.3594977092588\\
23.409090909091	2.03851611725483\\
23.5227272727275	7.21319451430441\\
23.6363636363637	10.2574640893986\\
23.75	12.3232754917954\\
23.8636363636365	13.8018457050268\\
23.9772727272727	14.8720281073956\\
24.0909090909092	15.6286636159955\\
24.2045454545457	16.1259051617722\\
24.318181818182	16.3950126906707\\
24.4318181818182	16.4524484742475\\
24.5454545454545	16.3035198695487\\
24.659090909091	15.9434669997372\\
24.7727272727275	15.3565000520001\\
24.8863636363637	14.5123412790102\\
25.0000000000002	13.35848841692\\
25.1136363636365	11.8032631855451\\
25.2272727272727	9.67488880700497\\
25.3409090909092	6.60205924154722\\
25.4545454545455	1.51889313037977\\
25.568181818182	-12.8896599044137\\
25.6818181818182	-2.71358351539975\\
25.7954545454547	4.41063213944926\\
25.909090909091	8.10169841338262\\
26.0227272727273	10.5035542045347\\
26.1363636363637	12.196061441441\\
26.25	13.4201521953399\\
26.3636363636365	14.2973189945604\\
26.477272727273	14.8952885496887\\
26.5909090909092	15.2532252093708\\
26.7045454545455	15.3928277627812\\
26.8181818181818	15.3234218407606\\
26.9318181818182	15.0439235676294\\
27.0454545454547	14.5426229187681\\
27.159090909091	13.794686938913\\
27.2727272727275	12.756147984829\\
27.3863636363637	11.3508240553516\\
27.5000000000002	9.43996872706652\\
27.6136363636365	6.7398972978837\\
27.7272727272727	2.52527316226279\\
27.8409090909092	-6.38774317356996\\
27.9545454545455	-8.38144770194769\\
28.068181818182	1.74886795257404\\
28.1818181818185	6.13707308002734\\
28.2954545454547	8.86253696487401\\
28.409090909091	10.7494517032786\\
28.5227272727273	12.1084540789577\\
28.6363636363637	13.0886736725724\\
28.7500000000002	13.7711644966907\\
28.8636363636365	14.2025107138338\\
28.977272727273	14.4091125397213\\
29.0909090909092	14.4037258360179\\
29.2045454545455	14.1882292974618\\
29.318181818182	13.7540131661285\\
29.4318181818182	13.0801704414986\\
29.5454545454547	12.1286173056846\\
29.659090909091	10.8334080484633\\
29.7727272727275	9.07654646369556\\
29.8863636363637	6.62547159758941\\
30	2.92729254509042\\
30.1136363636365	-3.99813909326786\\
30.2272727272727	-16.7058055598301\\
30.3409090909092	-0.798013572997302\\
30.4545454545457	4.34435981023731\\
30.568181818182	7.3809215672368\\
30.6818181818182	9.44469901384024\\
30.7954545454545	10.9225049601356\\
30.909090909091	11.9915219460657\\
31.0227272727275	12.7457002160009\\
31.1363636363637	13.2385422158496\\
31.2500000000002	13.500682317966\\
31.3636363636365	13.5478504875495\\
31.4772727272727	13.3843842985718\\
31.5909090909092	13.0041310128573\\
31.7045454545455	12.3891652731278\\
31.818181818182	11.5057116336117\\
31.9318181818182	10.2950719264176\\
32.0454545454547	8.65340035377598\\
32.159090909091	6.38122424559288\\
32.2727272727273	3.02743426392987\\
32.3863636363637	-2.83855155550256\\
32.5	-36.8279445870085\\
32.6136363636365	-3.26873577809538\\
32.727272727273	2.6984960832308\\
32.8409090909092	6.03539359847526\\
32.9545454545457	8.26112144530004\\
33.068181818182	9.84449260960044\\
33.1818181818182	10.9908740185285\\
33.2954545454547	11.8066325194621\\
33.409090909091	12.351723352816\\
33.5227272727275	12.6606186698682\\
33.6363636363637	12.7516624869896\\
33.7500000000002	12.631277932363\\
33.8636363636365	12.2953169569535\\
33.9772727272727	11.7281991963286\\
};

\addplot[area legend, draw=green, fill=green, draw opacity=0.1, fill opacity=0.1, forget plot]
table[row sep=crcr] {%
x	y\\
6	-20\\
14	-20\\
14	50\\
6	50\\
}--cycle;

\addplot[area legend, draw=red, fill=red, draw opacity=0.1, fill opacity=0.1, forget plot]
table[row sep=crcr] {%
x	y\\
9	-20\\
11	-20\\
11	50\\
9	50\\
}--cycle;
\end{axis}

\end{tikzpicture}%